\documentclass[twocolumn]{emulateapj}
\shorttitle{AGN Variability}
\shortauthors{Kelly et al.}

\begin{document}
  
  \title{Are the Variations in Quasar Optical Flux Driven by Thermal
  Fluctuations?}

  \author{Brandon C. Kelly\altaffilmark{1,2,3,4}, Jill Bechtold\altaffilmark{4}, 
    Aneta Siemiginowska\altaffilmark{3}}

  \altaffiltext{1}{bckelly@cfa.harvard.edu}
  \altaffiltext{2}{Hubble Fellow}
  \altaffiltext{3}{Harvard-Smithsonian Center for Astrophysics, 
      60 Garden St, Cambridge, MA 02138}
  \altaffiltext{4}{Department of Astronomy, University of Arizona, Tucson, AZ 85721}
  
  \begin{abstract}

    We analyze a sample of optical light curves for 100 quasars, 70 of
    which have black hole mass estimates. Our sample is the largest
    and broadest used yet for modeling quasar variability. The sources
    in our sample have $z < 2.8$, $10^{42} \lesssim \lambda
    L_{\lambda} (5100$\AA$) \lesssim 10^{46}$, and $10^6 \lesssim
    M_{BH} / M_{\odot} \lesssim 10^{10}$. We model the light curves as
    a continuous time stochastic process, providing a natural means of
    estimating the characteristic time scale and amplitude of quasar
    variations. We employ a Bayesian approach to estimate the
    characteristic time scale and amplitude of flux variations; our
    approach is not affected by biases introduced from discrete
    sampling effects. We find that the characteristic time scales
    stongly correlate with black hole mass and luminosity, and are
    consistent with disk orbital or thermal time scales. In addition,
    the amplitude of short time scale variations is significantly
    anti-correlated with black hole mass and luminosity. We interpret
    the optical flux fluctuations as resulting from thermal
    fluctuations that are driven by an underlying stochastic process,
    such as a turbulent magnetic field. In addition, the intranight
    variations in optical flux implied by our empirical model are
    $\lesssim 0.02$ mag, consistent with current microvariability
    observations of radio-quiet quasars. Our stochastic model is
    therefore able to unify both long and short time scale optical
    variations in radio-quiet quasars as resulting from the same
    underlying process, while radio-loud quasars have an additional
    variability component that operates on time scales $\lesssim 1$
    day.
    
  \end{abstract}
  
  
  \section{INTRODUCTION}

  \label{s-intro}

  It is widely accepted that the extraordinary activity associated
  with quasars\footnote[5]{Throughout this work we will use the terms
    quasar and AGN to refer generically to broad line AGNs. No
    luminosity difference between the two is assumed.} involves
  accretion onto a supermassive black hole, with the UV/optical
  emission arising from a geometrically thin, optically thick cold
  accretion disk. Aperiodic variability across all wavebands is
  ubiquitous in AGN, with the most rapid variations occuring in the
  X-rays \citep[for a review, see][]{ulrich97}. The source of quasar
  variability is unclear, and several models have been proposed for
  describing the optical variability of quasars, including accretion
  disk instabilities \citep[e.g.,][]{kawa98}, supernovae
  \citep[e.g.,][]{aretxaga97}, microlensing \citep{hawkins00}, and
  more general Poisson process models \citep[e.g.,][]{cid00}. However,
  recent results from reverberation mapping have shown that the broad
  emission lines respond to variations in the continuum emission after
  some time lag \citep[e.g.,][]{peter04}, implying that the continuum
  variations are dominated by processes intrinsic to the accretion
  disk. If the optical/UV variations are intrinsic to the accretion
  disk, then thermal fluctuations appear to be a natural choice for
  driving the optical/UV variations, as the optical/UV emission is
  thought to be thermal emission from the accretion disk. The fact
  that quasars become bluer as they brighten is consistent with a
  thermal origin \citep[e.g.,][]{giveon99,trevese01,geha03}. Moreover,
  the thermal timescale is sensitive to the disk
  viscosity. Variability is therefore a potentially important and
  powerful probe of the quasar central engine and accretion disk
  physics.

  A considerable amount of our interpretation and understanding of AGN
  accretion disks, and consequently their optical/UV emission, is
  based on the so-called $\alpha$-prescription \citep{shak73}. Within
  the standard $\alpha$ model, the viscosity, thought to be the source
  of the thermal disk emission and outward transfer of angular
  momentum, is parameterized as being proportional to the total
  pressure in the disk. Previous work on estimating $\alpha$ from
  quasar variability has found a value of $\alpha \sim 0.01$
  \citep[e.g.,][]{aneta89,collier01,starling04}. However, when the
  disk is dominated by radiation pressure, as is thought to be the
  case in the inner regions, an $\alpha$-disk is both
  thermally and viscously unstable \citep{shak76,lightman74}. In
  particular, for a radiation pressure dominated disk, the thermal
  instability is expected to grow exponentially on a time scale
  similar to the thermal time scale. For AGN, the thermal time scale
  is on the order of months to years, and in general there is no
  evidence for instabilities in the optical light curves of AGN which
  span $\gtrsim t_{th}$ \citep[e.g.,][]{collier01}. However, the
  optical light curves for a few sources may show evidence of
  instability on a thermal time scale \citep[e.g.,][]{czerny03,
    lub92}.

  If the stress is not proportional to the total pressure, but rather
  some other combination of the gas and radiation pressure, then the
  disk becomes more stable to thermal perturbations
  \citep[e.g.,][]{stella84,szusz90,merloni02,merloni03}. Alternatively,
  if part of the accretion energy is dissipated in a hot corona, then
  there is no longer any runaway heating in the disk, as additional
  cooling is provided by the corona \citep{svensson94}. For example, disk models
  with alternative prescriptions for $\alpha$, as well as the addition
  of a corona and jet, have been able to reproduce the variability of
  the microquasar GRS 1915+105 \citep{nayak00,janiuk00,janiuk02}. Previous work
  on 3-dimensional magneto-hydrodynamic (MHD) simulations of radiation
  pressure-dominated AGN accretion disks have not observed a thermal
  instability \citep[e.g.,][]{turner04,hirose08}. Moreover, the most
  promising physical mechanism behind the viscous torque is the
  magneto-rotational instability \citep[MRI,][]{balbus91,balbus98},
  and recent numerical and analytical work has suggested that the
  $\alpha$ prescription may be a poor representation for MRI-driven
  viscosity \citep{pessah08}. Variability studies can therefore lend
  observational insight into the inadequacies of the traditional
  $\alpha$-prescription, and possibly lead to a more accurate
  characterisation of the relationship between stress and pressure in
  the accretion disk.

  A successful model for quasar X-ray variability describes the X-ray
  variations on long time scales as being the result of perturbations
  in the accretion rate that occur outside of the X-ray emitting
  region \citep[e.g.,][]{lyub97,mayer06,janiuk07}. These accretion
  rate perturbations then travel inward, modulating the X-ray emitting
  region. It has been suggested that the origin of such perturbations
  is the result of a magnetic field randomly varying in time, and may
  be related to the appearance of an outflow \citep{king04}. If this
  model for the X-ray variability is correct, we would expect to also
  see variations in the optical luminosity, whose origin lies in the
  disk at radii farther from the central source. Therefore,
  understanding the origin of quasar optical variations will not only
  lead to a better understanding of accretion disk physics, but may
  also lead to a better understanding of the origin of quasar X-ray
  variability, possibly unifying the source of variability in the two
  bands.

  Quasars have also been observed to vary on time scales as short as
  hours \citep[so-called `microvariability' or `intranight
    variability',][]{gopal03,stalin04,stalin05,gupta05,carini07}. Microvariability
  is known to be stronger in radio-loud quasars, especially blazars
  \citep[e.g.,][]{gupta05}, while microvariability in radio-quiet
  quasars is usually not detected above the photometric uncertainty
  \citep[e.g.,][]{gupta05,carini07}. The enhanced level of
  microvariability in radio loud objects suggests that it is due to
  processes in a jet, while the physical cause of microvariability in
  radio quiet objects has remained a puzzle. Reprocessing of X-rays,
  disk instabilities, or a weak blazar component have been suggested
  \citep{czerny08}.

  There have been numerous previous investigations of quasar optical
  variability. However, because of the difficulty in obtaining high
  quality, well sampled light curves that cover a long time span, most
  previous work has involved ensemble studies of quasars, or analysis
  of simple correlations involving variability amplitude. The most
  well known result from previous work is a tendency for AGN to become
  less variable as their luminosity increases \citep[e.g.,][and
    references
    therein]{hook94,garcia99,giveon99,geha03,vanden04,devries05}. There
  have also been claims of a variability--redshift correlation
  \citep[e.g.,][]{crist90,cid96,crist96,trevese02}, although the sign
  of this correlation varies between studies \citep[e.g., see the list
    in][]{giveon99}. If real, the variability--redshift correlation is
  most likely caused by the fact that quasars are more variable at
  shorter wavelengths
  \citep[e.g.,][]{cutri85,diclemente96,helfand01,vanden04},
  corresponding to regions in the disk closer to the central black
  hole. Recently, a correlation between optical variability and black
  hole mass has been claimed \citep{wold07}. While many of these
  correlations are formally statistically significant, they often
  exhibit considerable scatter when one measures variability for
  individual objects.

  A few previous studies have employed spectral techniques, such as
  power spectra and structure functions, in the analysis of quasar
  optical light curves. From these studies it has been inferred that
  quasar optical light curves generally have variations of $\sim 10\%$
  on timescales of months, and that the power spectra of optical light
  curves is well described as $P(f) \propto 1 / f^2$
  \citep{giveon99,collier01,czerny03}; power spectra of this form are
  consistent with random walk, or more generally, autoregressive
  processes. In addition \citet{collier01} analyzed a sample of
  optical light curves from 8 low-$z$ Seyfert 1 galaxies. They found
  that the characteristic time scales of optical variations for the
  AGN in their sample are $\sim 10$--$100$ days and correlate with
  black hole mass, consistent with disk orbital or thermal time
  scales. \citet{czerny99} found evidence for a flattening of the
  optical power spectrum of NGC 5548 on time scales longer than $\sim
  100$ days. Both \citet{czerny99} and \citet{czerny03} also suggested
  that the long time scale optical variability was due to thermal
  fluctuations in the accretion disks of NGC 5548 and NGC 4151,
  respectively.

  Motivated by the potential in variability studies for increasing our
  understanding of the structure of quasar accretion disks, we have
  compiled a sample of well-sampled optical light curves from the
  literature. We directly model the quasar optical light curves as a
  stochastic process, in contrast to previous work based on more
  traditional Fourier (i.e., spectral) techniques. Our method allows
  us to describe quasar light curves with three free parameters: a
  characteristic time scale, amplitude of short time scale
  variability, and the mean value of the light curve. In addition, our
  method enables us to estimate the characteristic time scale of
  quasar variations without the windowing effects that can bias
  spectral approaches. Our sample consists of 100 AGN, 70 of which
  have black hole mass estimates. The sources in our sample have $z <
  2.8$, $10^{42} \lesssim \lambda L_{\lambda} (5100$\AA$) \lesssim
  10^{46}$, and $10^6 \lesssim M_{BH} / M_{\odot} \lesssim 10^{10}$,
  making this by far the largest sample yet used for this kind of
  study.

  In this work we adopt a cosmology based on the WMAP results
  \citep[$h=0.71, \Omega_m=0.27, \Omega_{\Lambda}=0.73$,][]{wmap}.

  \section{DATA}

  \label{s-data}

  In this work we analyze a sample of 100 quasar optical light curves,
  compiled from the literature. Our sample consists of 55 AGN from the
  \emph{MACHO} survey \citep{geha03}, 37 Palomar Green (PG) quasars
  from the sample of \citet{giveon99}, and 8 Seyfert galaxies from the
  AGN
  Watch\footnote[1]{http://www.astronomy.ohio-state.edu/\~{}agnwatch/}
  database. We were able to obtain black hole mass estimates for 71 of
  the AGN, where $M_{BH}$ has been estimated for 20 of them from
  reverberation mapping \citep{peter04}, and $M_{BH}$ is estimated for
  the remaining 51 AGN from the broad emission lines using standard
  scaling relationships \citep[e.g.,][]{vest06}. Our sample is
  summarized in Table \ref{t-sample}.

\begin{deluxetable*}{ccccccc}
\tablecaption{Quasars Analyzed in this Work\label{t-sample}}
\tablewidth{0pt}
\tablehead{
\colhead{RA} 
& \colhead{DEC}
& \colhead{$z$} 
& \colhead{$\log \lambda L_{\lambda} (5100$\AA$)$} 
& \colhead{$\log M_{BH}$} 
& \colhead{Err. in $\log M_{BH}$\tablenotemark{a}} 
& \colhead{Ref.\tablenotemark{b}} \\
\colhead{J2000} & \colhead{J2000} & \colhead{} & 
\colhead{${\rm erg\ s^{-1}}$} & \colhead{$M_{\odot}$} & \colhead{} &
}
\startdata
 00 29 13.7 & +13 16 03.9 & 0.142 & 45.02 & 8.59 & 0.10 & 1  \\
 00 47 15.8 & -72 41 12.2 & 0.530 & 44.34 & \nodata & \nodata & 2  \\
 00 49 34.4 & -72 13 09.0 & 0.610 & 44.56 & \nodata & \nodata & 2  \\
 00 51 16.9 & -72 16 51.1 & 0.330 & 44.00 & \nodata & \nodata & 2  \\
 00 54 52.1 & +25 25 39.0 & 0.155 & 44.96 & 8.56 & 0.08 & 1  \\
 00 55 34.2 & -72 28 30.0 & 1.660 & 45.69 & \nodata & \nodata & 2  \\
 00 55 59.6 & -72 52 45.1 & 0.170 & 43.74 & \nodata & \nodata & 2  \\
 01 01 27.8 & -72 46 14.4 & 1.720 & 45.52 & \nodata & \nodata & 2  \\
 01 02 14.4 & -73 16 26.8 & 1.640 & 45.47 & \nodata & \nodata & 2  \\
 01 02 34.7 & -72 54 22.2 & 0.220 & 43.93 & \nodata & \nodata & 2  \\
 01 07 21.7 & -72 48 45.8 & 0.280 & 44.14 & \nodata & \nodata & 2  \\
 04 46 11.0 & -72 05 09.0 & 0.900 & 45.08 & \nodata & \nodata & 2  \\
 04 53 56.6 & -69 40 36.0 & 0.460 & 45.19 & 8.41 & 0.45 & 2  \\
 04 56 14.2 & -67 39 10.8 & 2.220 & 45.22 & \nodata & \nodata & 2  \\
 05 00 17.6 & -69 32 16.3 & 1.050 & 44.93 & \nodata & \nodata & 2
\enddata

\tablecomments{Table \ref{t-sample} is published in its entirety in the
  electronic edition of the {\it Astrophysical Journal}. A portion is
  shown here for guidance regarding its form and content.}

\tablenotetext{a}{1$\sigma$ uncertainty on $\log M_{BH}$.}
\tablenotetext{b}{Reference for the optical light curve data.}

\tablerefs{
(1) Giveon et al. 1999
(2) Geha et al. 2003
(3) Stirpe et al. 1994
(4) Peterson et al. 2000
(5) Kaspi et al. 1996
(6) Santos-Lleo et al. 2001
(7) Peterson et al. 2002
(8) Carone et al. 1997
(9) Shemmer et al. 2001
(10) Collier et al. 1998
}
\end{deluxetable*}

  \subsection{Macho Quasars from Geha et al.(2003)}

  \label{s-macho}

  We collected $R$ band light curves from AGN selected by the
  \emph{MACHO} survey based on their variability
  \citep{alcock97,alcock99}. The motivation for the \emph{MACHO}
  survey was to study Galactic microlensing events behind the
  Magellanic Clouds. However, the survey was also able to select
  quasars via their variability, confirmed via spectroscopic
  follow-up, producing 59 quasars with well-sampled light curves over
  a broad range in redshift ($0.1 \lesssim z \lesssim 2.8$)
  \citep{geha03}. The quasar light curves span $\sim 7.5$ years, and
  have a typical sampling interval of $\sim 2$--$10$ days, although
  much longer gaps exist for some light curves. In general, the
  \emph{MACHO} light curves are the highest quality in our sample,
  often being frequently and regularly sampled. However, we note that
  because the \emph{MACHO} sources were selected based on their
  variability, this sample is biased toward more variable objects.
  See \citet{geha03} for more details of the \emph{MACHO} quasar
  catalogue.

  Spectra for the \emph{MACHO} quasars are presented in
  \citet{geha03}, and spectra for 27 of these quasars were kindly
  provided to us by Marla Geha. We calculated black hole mass
  estimates from the source luminosity and the $FWHM$ of the H$\beta$,
  Mg II, or C IV broad emission line using standard scaling
  relationships \citep[e.g.,][Vestergaard 2008]{vest06}. However, the
  spectra are not flux-calibrated, so the luminosities at $1350,
  3000,$ and $5100$\AA\ were estimated from the photometric data,
  assuming a power-law continuum with $\alpha = 0.5, f_{\nu} \propto
  \nu^{-\alpha}$ \citep{rich01}. Typical uncertainties on the broad
  line mass estimates are $\sim 0.4$ dex \citep[e.g.,][]{vest06}. When
  combined with the measurement error in the $FWHM$ measurements, and
  the uncertainty due to the intrinsic scatter in quasar spectral
  slopes \citep[$\sim 0.3$,][]{rich01}, our typical adopted
  uncertainty in $M_{BH}$ was $\sim 0.45$ dex for the \emph{MACHO}
  quasars.

  \subsection{PG Quasars from Giveon et al.(1999)}

  \label{s-giveon}

  \citet{giveon99} report $B$ and $R$ photometric light curves for 42
  PG quasars spanning $\sim 7$ years with a typical sampling interval
  of $\sim 40$ days. These quasars are all bright ($B < 16$ mag) and
  nearby ($z < 0.4$). \citet{peter04} report estimates of $M_{BH}$ and
  $L_{\lambda} (5100$\AA$)$ calculated from reverberation mapping for
  12 of these quasars. Estimates of $M_{BH}$ and
  $L_{\lambda}(5100$\AA$)$ were taken from \citet{vest06} for the
  remaining \citet{giveon99} quasars. See \citet{giveon99} for further
  details.

  \subsection{Seyfert Galaxies from AGN Watch Database}

  \label{s-agnwatch}

  The light curves for the remaining 8 AGN in our sample are from the
  AGN watch project. The optical light curves for the Seyfert Galaxies
  AKN 564, Fairall 9, MRK 279, MRK 509, NGC 3783, NGC 4051, NGC 4151,
  NGC 5548, and NGC 7469 were taken from the AGN watch website. In
  general, we used the light curves at 5100\AA, with the exception of
  AKN 564, for which we used the $R$-band light curve. We excluded
  3C390.3 from our analysis because it is a classified as an
  optically-violent variable, and thus may experience a different
  variability mechanism than the other AGN in our sample. Black hole
  masses and 5100\AA\ luminosities were taken from
  \citet{peter04}. Further details may be found in the references
  listed in Table \ref{t-sample}. Although the AGN Watch sample is
  small compared to the other two, these sources are particularly
  important in our analysis because they help to anchor the low-$L$
  and low-$M_{BH}$ end of the correlations analyzed in
  \S~\ref{s-results}.

  \section{THE STATISTICAL MODEL: CONTINUOUS AUTOREGRESSIVE PROCESS}

  \label{s-car1} 

\begin{figure}
  \begin{center}
    \scalebox{0.8}{\rotatebox{90}{\plotone{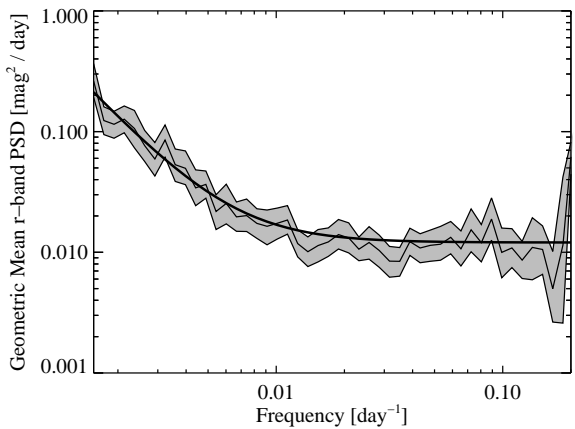}}}
    \caption{The geometric mean power spectrum (solid noisy line) for
    the $R$-band light curves of the \emph{MACHO} quasars, along with
    $90\%$ confidence region (shaded region). The thick solid line is
    a power spectrum of the form $P(f) \propto 1 / f^2$ with an
    additive measurement error contribution. The optical light curves
    for the \emph{MACHO} quasars are well described by a $1 / f^2$
    power spectrum, consistent with other samples of quasars. Power
    spectra of the form $1 / f^2$ are suggestive of random walk and
    related stochastic processes.
    \label{f-macho_pspec}}
  \end{center}
\end{figure}

  The previous analysis of \citet{giveon99} and \citet{collier01}
  suggests that the optical variability can be well-described by a
  power-law power spectrum with slope $\sim 2$. The power spectra of
  the \emph{MACHO} quasars have not been analyzed for individual
  objects, although \citet{hawkins07} investigated the power spectra
  of the ensemble of objects. In Figure \ref{f-macho_pspec} we show
  the geometric mean $R$-band power spectra for the \emph{MACHO}
  quasars, along with the 90\% confidence region on the geometric
  mean. The power spectra for the \emph{MACHO} sources are well
  described by a power law of slope $\sim 2$, consistent with previous
  work. The lack of any peaks in the power spectra, as well as the
  aperiodic and noisy appearance of quasar light curves, suggests that
  quasar light curves are stochastic or chaotic in nature. 

  In this work we model quasar light curves as a {\it continuous time first
  order autoregressive process} (CAR(1)). Power spectra of the form
  $P(f) \propto 1 / f^2$ are consistent with a first order
  autoregressive (AR(1)) process. We model this stochastic process in
  continuous time, both because the actual physical processes in the
  accretion disk are continuous, and because doing so allows a natural
  way of handling the irregular sampling of our light
  curves. Moreover, this process is well-studied, only has three
  parameters, and provides a natural and consistent way of estimating
  a characteristic time scale and variance of quasar light curves.

  The CAR(1) process is described by the following stochastic
  differential equation\footnote[2]{Strictly speaking, the stochastic
    differential equation is complicated by the fact that white noise
    does not exist as a derivative in the usual sense. However, we
    ignore the mathematical technicalities for ease of interpretation
    of Equation (\ref{eq-car1})} \citep[e.g.,][]{brockwell02}:
  \begin{equation}
    dX(t) = -\frac{1}{\tau} X(t) dt + \sigma \sqrt{dt} \epsilon(t) + b\ dt,
    \ \ \ \tau, \sigma, t > 0.
    \label{eq-car1}
  \end{equation}
  Here, $\tau$ is called the `relaxation time' of the process $X(t)$,
  and $\epsilon(t)$ is a white noise process with zero mean and
  variance equal to one. Within the context of this work, $X(t)$ is
  the quasar flux. We assume that the white noise process is
  also Gaussian. The mean value of $X(t)$ is $b\tau$ and the variance
  is $\tau \sigma^2 / 2$. Further details on the CAR(1) process are
  described in the Appendix.

  The relaxation time, $\tau$, can be interpreted as the time required
  for the time series to become roughly uncorrelated, and $\sigma$ can
  be interpreted as describing the variability of the time series on
  time scales short compared to $\tau$. Within the context of this
  work, $X(t)$ is the quasar light curve. It is tempting to associate
  $\tau$ with a characteristic time scale, such as the time required
  for diffusion to smooth out local accretion rate perturbations, and
  $\sigma$ to represent the variability resulting from local random
  deviations in the accretion disk structure, such as caused by
  turbulence and other random magneto-hydrodynamic (MHD) effects.

  The power spectrum of a CAR(1) process is
  \begin{equation}
    P_{X}(f) = \frac{2 \sigma^2 \tau^2}{1 + (2 \pi \tau f)^2}.
    \label{eq-pspec}
  \end{equation}
  From Equation (\ref{eq-pspec}) we infer that there are two important
  regimes for $P_{X}(f)$: $P_X (f) \propto 1 / f^2$ for $f \gtrsim
  (2\pi\tau)^{-1}$ and $P_X (f) \propto {\rm constant}$ for $f \lesssim (2 \pi
  \tau)^{-1}$. Therefore, the CAR(1) process has a power spectrum that
  falls off as $1 / f^2$ at time scales short compared to the
  relaxation time, and flattens to white noise at time scales long
  compared to the relaxation time. Because `characteristic' time
  scales of quasar light curves are often defined by a break in the
  power spectrum, this is an additional justification of associating
  $\tau$ with a characteristic time scale. In addition, because the
  power spectra of quasar optical light curves are well described by
  $P_X(f) \propto 1 / f^2$, it suggests that a CAR(1) process should
  provide a good description of the light curves, with $\tau$ being on
  the order of the length of the light curves or longer.

  To illustrate the CAR(1) process, we simulate four CAR(1) light
  curves. The light curves were simulated by first simulating a random
  variable from a normal distribution with mean $\tau b$ and variance
  $\tau \sigma^2 / 2$; note that this is the mean and variance of the
  CAR(1) process. Then, from this random initial value, we simulated
  the rest of the light curve using Equations (\ref{eq-cexpect}) and
  (\ref{eq-cvar}) in the Appendix. These simulated light curves span a
  length of 7 yrs and are sampled every 5 days. The simulated light
  curves span a period in time similar to the quasar light curves
  analyzed in this work, but are better sampled than most of the
  quasar light curves. Three characteristic time
  scales of interest for quasars are the light crossing time, the gas
  orbital time scale, and the accretion disk thermal time scale. These
  time scales are
  \begin{eqnarray}
    t_{lc} & = & 1.1 \times \left(\frac{M_{BH}}{10^8 M_{\odot}}\right) 
                 \left(\frac{R}{100 R_S}\right)\ {\rm days} \label{eq-tlc} \\
    t_{orb} & = & 104 \times \left(\frac{M_{BH}}{10^8 M_{\odot}}\right) 
                 \left(\frac{R}{100 R_S}\right)^{3/2}\ {\rm days} \label{eq-torb} \\
    t_{th} & = & 4.6 \times \left(\frac{\alpha}{0.01}\right)^{-1} \left(\frac{M_{BH}}{10^8 M_{\odot}} \right)
                 \left(\frac{R}{100 R_S}\right)^{3/2}\ {\rm yrs} \label{eq-tth},
  \end{eqnarray}
  where $M_{BH}$ is the mass of the black hole, $R$ is the emission
  distance from the central black hole, $R_S = 2 GM_{BH} / c^2$ is the
  Schwarzschild radius, and $\alpha$ is the standard disk viscosity
  parameter. For the simulated quasar light curves, we use $M_{BH} =
  10^8 M_{\odot}, \alpha = 0.01,$ and $R = 100 R_S$, and set $\tau$
  equal to each of these three time scales. In addition, we use $b = 0$
  and $\sigma = 1$. Note that the assumed value of $\alpha$ only
  affects the thermal time scale, and a higher value of $\alpha$
  results in a shorter time scale.

\begin{figure}
  \begin{center}
    \scalebox{1.0}{\plotone{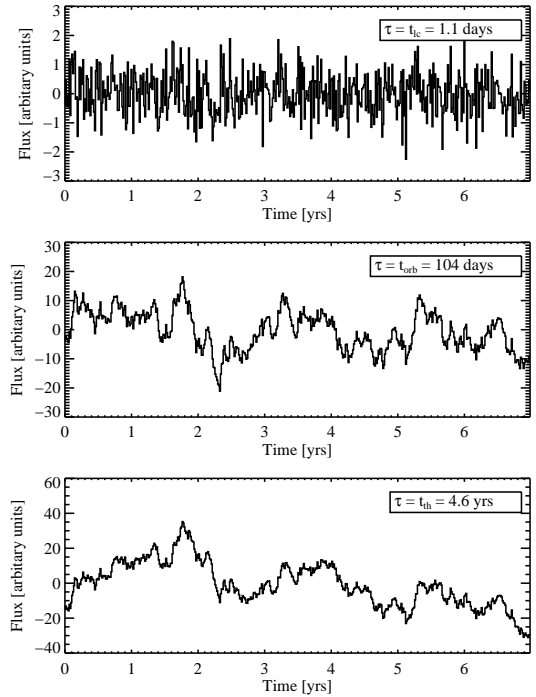}}
    \caption{Light curves simulated from a CAR(1) process for three
    different characteristic time scales, assuming typical parameters
    for quasars ($M_{BH} = 10^8 M_{\odot}, R_s = 100, \alpha = 0.01$,
    see Eq.[\ref{eq-tlc}]--[\ref{eq-tth}]). From top to 
    bottom, these are the light crossing time, $\tau = 1.1$ days, the
    disk orbital time 
    scale, $\tau = 104$ days, and the disk thermal time scale, $\tau =
    4.6$ yrs. The stochastic nature of the CAR(1) process is apparent,
    and the light curve exhibits more variability on longer time
    scales as the characteristic time scale increases.
    \label{f-car1_simlc}}
  \end{center}
\end{figure}

\begin{figure}
  \begin{center}
    \scalebox{1.0}{\plotone{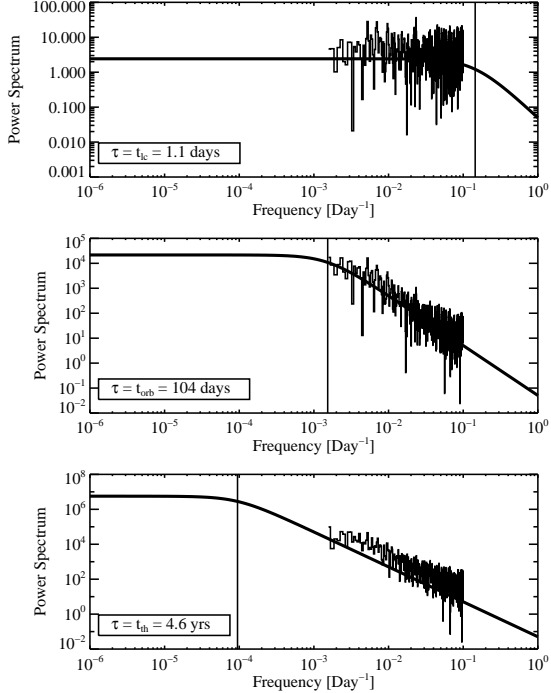}}
    \caption{Power spectra for the simulated CAR(1) light curves shown
      in Figure \ref{f-car1_simlc}. The actual power spectra are shown
      with a solid line, and the empirical power spectra estimated
      directly from the light curves are the noisy curves. The power
      spectra are flat on the `white noise' part of the curve,
      corresponding to frequencies $f \lesssim (2\pi\tau)^{-1}$, and
      fall off as $1 / f^2$ on the `red noise' part of the curve, $f
      \gtrsim (2\pi\tau)^{-1}$. As $\tau$ increases, the break in the
      power spectra, marked with a vertical line, shifts toward
      smaller frequencies. For the CAR(1) process with $\tau =
      t_{th}$, red noise leak biases the power spectrum estimated
      directly from the simulated light curve.
    \label{f-car1_pspec}}
  \end{center}
\end{figure}

  The simulated light curves are shown in Figure \ref{f-car1_simlc},
  and their corresponding power spectra are shown in Figure
  \ref{f-car1_pspec}. The increased amount of variation on long time
  scales with increasing $\tau$ is apparent. In addition, because
  $t_{lc}$ is shorter than the time sampling, the first simulated
  light curve is only sampling frequencies on the flat part of the
  power spectrum, giving it the appearance of white noise. In
  contrast, the two simulated light curves with the longest time
  scales are sampled on the $1 / f^2$ part of the power spectrum,
  giving them more of a `red noise' appearance. In addition, `red
  noise' leak affects the estimated power spectrum of the light curve
  with $\tau = 4.6$ yrs, evidenced by the constant offset between the
  true power spectrum and the estimated one. Red noise leak occurs
  when power from time scales longer than the span of the time series
  `leaks' into the shorter time scales, biasing the power spectrum
  when estimated as the modulus of the discrete Fourier transform
  \citep[e.g.,][]{klis97}.

  \subsection{Estimating the Parameters of a CAR(1) Process}

  \label{s-estimate}

\begin{figure*}
  \begin{center}
    \includegraphics[scale=0.33]{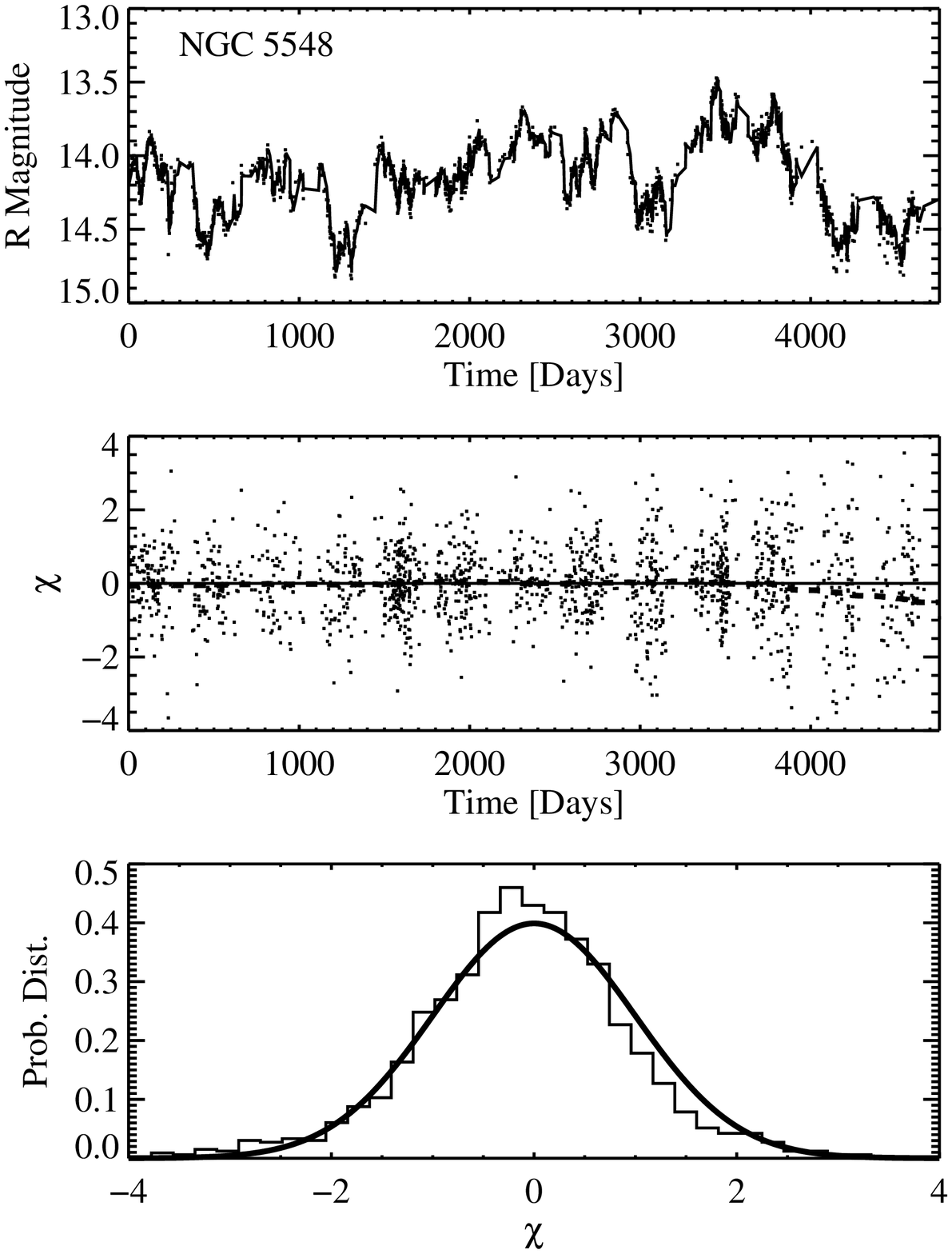}
    \includegraphics[scale=0.33]{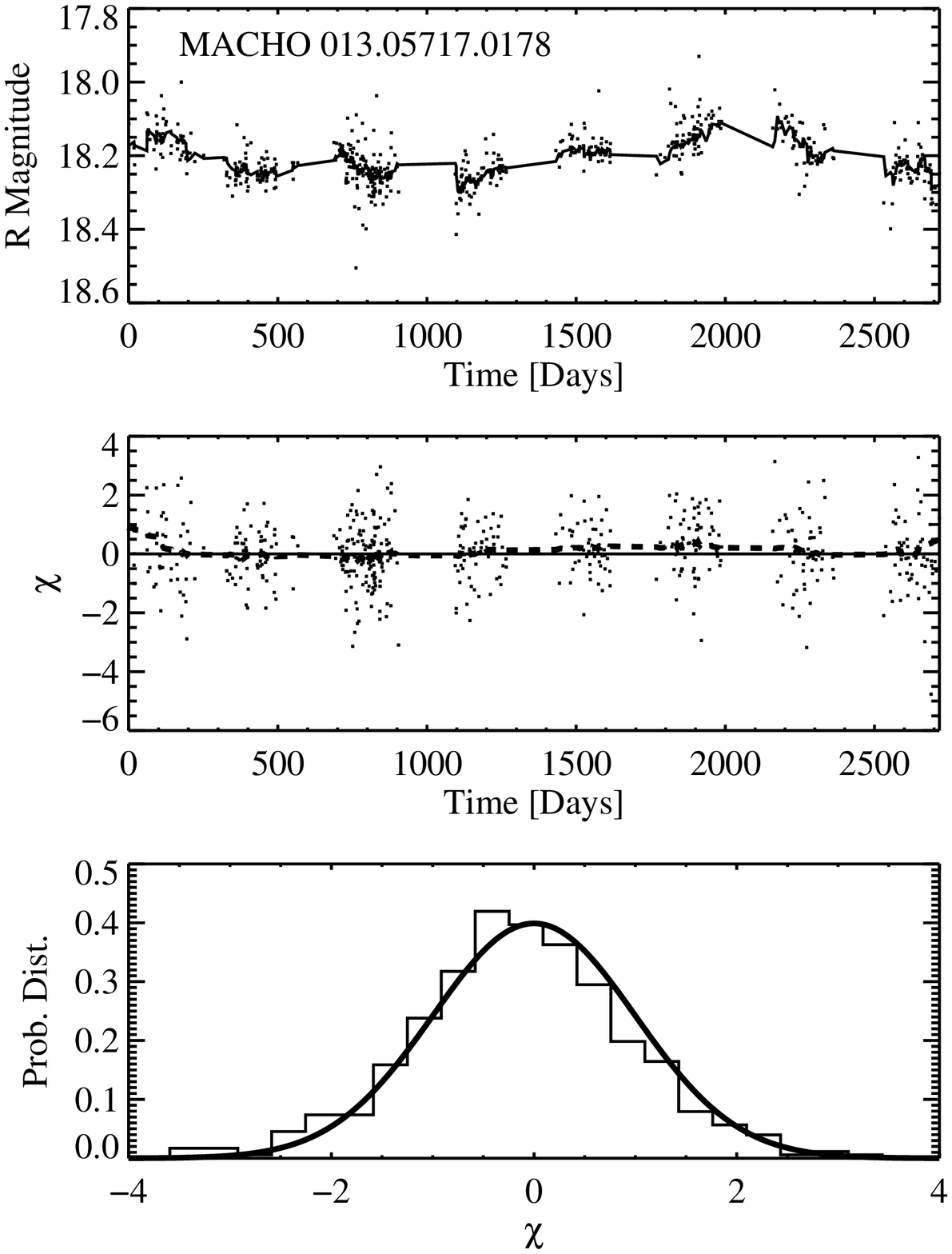}
    \includegraphics[scale=0.33]{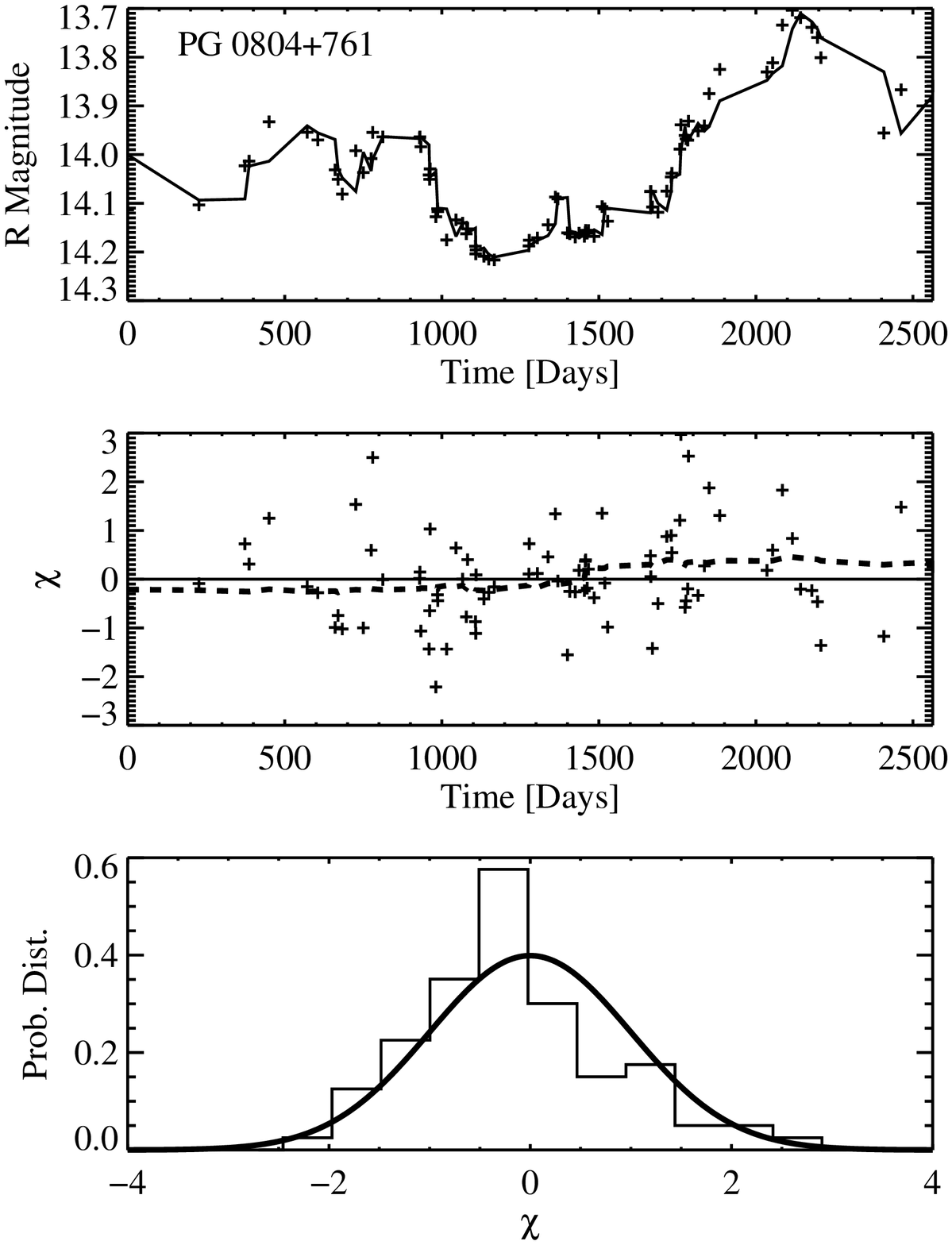}
    \caption{Light curves and best fit CAR(1) processes for NGC 5548
    (left), MACHO source 013.05717.0178 (middle), and PG 0804+761
    (right). The top panels show the light curves (data points) along
    with the best fit CAR(1) process (solid line), the middle panels
    show the standardized residuals (Eq.[\ref{eq-resids}]) as a
    function of time (data points) and a moving average estimate
    (dashed line), and the bottom panels compare a histogram of the
    standardized residuals with the expected standard normal
    distribution. The AGN light curves are well described by a CAR(1)
    process with characteristic time scales $\tau = 214, 6026,$ and $1148$
    days for NGC 5548, MACHO 013.05717.0178, and PG 0804+761, respectively.
    \label{f-lcfits}}
  \end{center}
\end{figure*}

  The parameters for a CAR(1) process are commonly estimated by
  maximum-likelihood directly from the observed time series. This is
  an advantage over non-parameteric approaches, such as the discrete
  power spectrum or the structure function. The observed power
  spectrum and structure function can both suffer from windowing
  effects caused by the finite duration and sampling of the light
  curve, whereby power from high frequencies can leak to low
  frequencies (aliasing), and power at low frequencies can leak to
  high frequencies (e.g., red noise leak). For ground based optical
  observations, an additional complication is the periodicity enforced
  in the sampling caused by the Earth's rotation around the sun, as
  objects are only observable during certain times of the year. For
  example, this periodic sampling can be seen in the light curve for
  the \emph{MACHO} source shown in Figure \ref{f-lcfits}. All of these
  effects can bias the power spectrum or structure function when
  estimated directly from the light curve in a non-parameteric
  fashion. Similarly, these effects can bias parameter estimates when
  fitting a model to the observed power spectrum or structure
  function.

  In contrast, estimating a `characteristic' time scale and variance
  directly from the observed time series, instead of from the observed
  power spectrum or structure function, has the advantage of being
  free of windowing effects, giving unbiased estimates of $\tau$ and
  $\sigma^2$. Of course, this requires one to assume a parameteric
  model for the time series (or power spectrum), but as we will show
  below the CAR(1) process provides a good description of most of the
  AGN light curves analyzed in this work. Furthermore, higher-order
  terms can be added to Equation (\ref{eq-car1}) to allow additional
  flexibility \citep[e.g.,][]{brockwell02}, but this is beyond the
  scope of the current work.

  When the light curve is measured with error, the likelihood function
  can be calculated using a `state-space' representation of the time
  series \citep[e.g.,][]{brockwell02}. Denoting the measured fluxes as
  $x_1,\ldots,x_n$, observed at times $t_1,\ldots,t_n$ with
  measurement error variances $\sigma_1^2, \ldots, \sigma_n^2$, the
  likelihood function, $p(x_1,\ldots,x_n|b,\sigma,\tau)$, is a product
  of Gaussian functions:
  \begin{eqnarray}
    p(x_1,\ldots,x_n|b,\sigma,\tau) & = & \prod_{i=1}^n \left[2\pi
      (\Omega_i + \sigma_i^2)\right]^{-1/2} \nonumber \\
      & & \times \exp\left\{ -\frac{1}{2} \frac{(\hat{x}_i - x^*_i)^2}{\Omega_i + \sigma^2_i} \right\}
      \label{eq-lik} \\
    x^*_i & = & x_i - b\tau \label{eq-xstar} \\
    \hat{x}_0 & = & 0 \label{eq-xhat0} \\
    \Omega_0 & = & \frac{\tau \sigma^2}{2} \\
    \hat{x}_{i} & = & a_{i} \hat{x}_{i-1} \nonumber \\ 
     & + & \frac{a_{i} \Omega_{i-1}}{\Omega_{i-1} + \sigma^2_{i-1}}
      \left(x^*_{i-1} - \hat{x}_{i-1}\right) \label{eq-xhati} \\
    \Omega_{i} & = & \Omega_0 \left(1 - a_i^2\right) \nonumber \\ 
      & + & a^2_i \Omega_{i-1} \left(1 - \frac{\Omega_{i-1}}{\Omega_{i-1} + \sigma^2_{i-1}} \right)
      \label{eq-omegai} \\
    a_i & = & e^{-(t_i - t_{i-1}) / \tau} \label{eq-acoef}.
  \end{eqnarray}
  The maximum-likelihood estimate is then found by maximizing Equation
  (\ref{eq-lik}) with respect to $b, \tau,$ and $\sigma$.

  In this work we employ a Bayesian approach in order to directly
  compute the probability distribution of $b, \tau,$ and $\sigma$,
  given our observed light curves. The probability distribution of the
  parameters, given the observed data (i.e., the \emph{posterior}
  distribution), is calculated as the product of the likelihood
  function with a prior probability distribution. In this work we
  assume a uniform prior on $b$ and $\sigma$. For deriving a prior on
  $\tau$, we note that when the data are regularly sampled, the CAR(1)
  process reduces to the AR(1) process described by Equation
  (\ref{eq-ar1}) in the Appendix with $\alpha_{AR} = e^{-\Delta
    t/\tau}$, where $\Delta t$ is the time sampling interval. In
  this work we assume a uniform prior on $\alpha_{AR}$. For an AR(1)
  process, $\alpha_{AR}$ gives the correlation between $x_i$ and
  $x_{i-1}$. Therefore, we consider it reasonable to assume that any
  value of $\alpha_{AR}$ is \emph{a priori} likely, i.e., we do not
  assume anything \emph{a priori} about the correlations between
  subsequent data points, and therefore we assume a uniform prior on
  $\alpha_{AR}$ from 0 to 1. This prior is non-informative in the
  sense that all of the information on $\alpha_{AR}$ comes from the
  data.

  The accuracy of the fit can be assessed by comparing the residuals
  of the light curve with the values expected under the assumption of
  a CAR(1) process. Equation (\ref{eq-lik}) implies that if the CAR(1)
  process provides a good model of the observed data, then the
  residuals should be uncorrelated and follow a normal distribution:
  \begin{equation}
    \chi \equiv \frac{x^*_i - \hat{x}_i}{\sqrt{\Omega_i + \sigma^2_i}} \sim N(0,1).
    \label{eq-resids}
  \end{equation}
  Here, the notation $\chi \sim N(0,1)$ means that $\chi$ is distributed
  according to a normal distribution with mean equal to zero and
  variance equal to one. The goodness of fit can then be assessed by
  inspecting a plot of the residuals with time to ensure that they are
  uncorrelated, and by comparing a histogram of the residuals with the
  expected standard normal distribution.

  \subsection{Fitting the Quasar Light Curves}

  \label{s-fits}

  In this work we model the logarithm of the flux as following a
  CAR(1) process, or equivalently the apparent magnitudes. We do this
  because the assumption of a Gaussian white noise process in Equation
  (\ref{eq-car1}) produces both positive and negative values of
  $X(t)$, while flux is a strictly positive quantity. The logarithm
  function maps a strictly positive quantity to the interval
  $(-\infty, \infty)$, and therefore Equation (\ref{eq-car1}) is
  likely to be a better description of the light curve for the source
  apparent magnitude, as opposed to the source flux. \citet{uttley05}
  have also argued for describing accreting black holes as a Gaussian
  stochastic process in the logarithm of the flux.

  An additional correction is needed to correct for cosmological time
  dilation. Because the quasars are fit in the observed frame, and
  time scales decrease as $(1 + z)$, spurious correlations may arise
  if one does not correct to the quasar rest frame. This is
  particularly problematic when dealing with flux limited samples,
  which create an artificial correlation between $z$ and both
  luminosity and $M_{BH}$. Noting that $dt_{obs} = (1 + z) dt_{rest}$,
  Equation (\ref{eq-car1}) can be expressed in the forms
  \begin{eqnarray}
    dX(t) & = & -\frac{1}{\tau_{obs}} X(t) dt_{obs} + \sigma_{obs} \sqrt{dt_{obs}} \epsilon(t) + 
                b_{obs}\ dt_{obs} \label{eq-car1obs} \\
     & = & -\frac{1+z}{\tau_{obs}} X(t) dt_{rest} + \sigma_{obs}
     \sqrt{(1+z)dt_{rest}} \epsilon(t) \nonumber \\
     & + & (1+z) b_{obs}\ dt_{rest} \label{eq-car1rest1} \\
     & = & -\frac{1}{\tau_{rest}} X(t) dt_{rest} +
     \sigma_{rest}\sqrt{dt_{rest}} \epsilon(t) \nonumber \\
     & + & b_{rest}\ dt_{rest} 
    \label{eq-car1rest}.
  \end{eqnarray}
  From Equations (\ref{eq-car1obs})--(\ref{eq-car1rest}) it is
  apparent that the observed and rest frame parameters are related as
  \begin{eqnarray}
    \tau_{rest} & = & (1 + z)^{-1} \tau_{obs} \label{eq-taurest} \\
    \sigma_{rest} & = & (1 + z)^{1/2} \sigma_{obs} \label{eq-sigrest} \\
    b_{rest} & = & (1+z) b_{obs} \label{eq-brest}.
  \end{eqnarray}
  Noting that the mean of the CAR(1) process is $b\tau$, and that the
  variance is $\tau \sigma^2 / 2$, Equations
  (\ref{eq-taurest})--(\ref{eq-brest}) imply that the mean and
  variance of a CAR(1) process are unaffected by cosmological time
  dilation. However, the variance observed from a light curve with a
  finite duration and sampling is still affected by time dilation, as
  the observed variance over a time interval $\Delta t$ is the
  integral of Equation (\ref{eq-pspec}) over that time interval. In
  what follows, the quantities $\tau, \sigma,$ and $b$ will always
  refer to the quasar rest frame quantities, unless specified
  otherwise.

  Random draws of $b, \tau,$ and $\sigma$ from the posterior
  probability distribution are obtained using a Metropolis-Hastings
  algorithm. We calculated an estimate of each parameter as the median
  of the posterior, and the posterior medians, standard deviations,
  and confidence intervals are computed using these random draws. In
  general the posterior median values were not significantly different
  from a maximum likelihood fit. The goodness of fit for the light
  curves was determined by examining a histogram of the residuals and
  a plot of the residuals against time, as described in
  \S~\ref{s-estimate}. Occasionally outlying values of the flux are
  present for the light curves with more data points, possibly due to
  unidentified systematic error. These outlying data points were
  removed and the light curves were refit. In Figure \ref{f-lcfits} we
  show the light curve and best fit CAR(1) model for the most densely
  sampled object in our sample, NGC 5548, for a representative light
  curve from the \emph{MACHO} sample, and for a representative light
  curve from the PG sample of \citet{giveon99}. In general, the CAR(1)
  model provided a good fit to the quasar light curves analyzed in
  this work, and we only flagged 9 out of 109 light curves as having a
  bad fit. The best fit CAR(1) parameters, along with their
  uncertainties, for the 100 AGN in our sample are listed in Table
  \ref{t-car1}. The 9 objects for which the fit was deemed
  unacceptable are not used in the regression analysis.

\begin{deluxetable*}{cccccc}
\tablewidth{0pt}
\tablecaption{Results from CAR(1) Process Fits to Quasar Light Curves\label{t-car1}}
\tablehead{
\colhead{RA}
& \colhead{DEC}
& \colhead{$\log \tau$\tablenotemark{a}} 
& \colhead{Conf. Int. for $\log \tau$} 
& \colhead{$\log \sigma$\tablenotemark{b}} 
& \colhead{Err. in $\log \sigma$} \\
\colhead{J2000} & \colhead{J2000} & \colhead{day} & \colhead{$95\%$, day} & 
\colhead{${\rm mag\ day^{1/2}}$} & 
}
\startdata
 00 29 13.7 & +13 16 03.9 & 3.44 & [2.59,6.17] & -1.97 & 0.05  \\
 00 47 15.8 & -72 41 12.2 & \nodata\tablenotemark{c} & \nodata & \nodata & \nodata  \\
 00 49 34.4 & -72 13 09.0 & 2.97 & [2.33,5.50] & -1.95 & 0.05  \\
 00 51 16.9 & -72 16 51.1 & 2.10 & [1.84,2.70] & -1.60 & 0.03  \\
 00 54 52.1 & +25 25 39.0 & 2.71 & [2.18,4.82] & -1.81 & 0.05  \\
 00 55 34.2 & -72 28 30.0 & 2.56 & [2.08,4.14] & -2.31 & 0.06  \\
 00 55 59.6 & -72 52 45.1 & 3.07 & [2.38,6.86] & -1.85 & 0.02  \\
 01 01 27.8 & -72 46 14.4 & 3.04 & [2.39,5.22] & -2.04 & 0.04  \\
 01 02 14.4 & -73 16 26.8 & 2.75 & [2.22,5.28] & -2.23 & 0.05  \\
 01 02 34.7 & -72 54 22.2 & 2.41 & [2.04,3.96] & -1.80 & 0.03  \\
 01 07 21.7 & -72 48 45.8 & 2.34 & [2.00,3.74] & -1.89 & 0.03  \\
 04 46 11.0 & -72 05 09.0 & 3.03 & [2.41,5.34] & -2.20 & 0.05  \\
 04 53 56.6 & -69 40 36.0 & 1.78 & [1.60,2.07] & -1.93 & 0.02  \\
 04 56 14.2 & -67 39 10.8 & 3.26 & [2.55,5.53] & -2.17 & 0.06  \\
 05 00 17.6 & -69 32 16.3 & \nodata\tablenotemark{c} & \nodata & \nodata & \nodata  \\
\enddata

\tablecomments{Table \ref{t-car1} is published in its entirety in the
  electronic edition of the {\it Astrophysical Journal}. A portion is
  shown here for guidance regarding its form and content.}

\tablenotetext{a}{The logarithm of the characteristic time scale of the
  quasar lightcurve, in days.}  
\tablenotetext{b}{The logarithm of the standard deviation in the input
  process to Equation (\ref{eq-car1}). The standard deviation of
  quasar flux variations on time scales of 1 day is expected to be
  equal to $\sigma$.}
\tablenotetext{c}{The CAR(1) process provided a poor fit to these
  lightcurves, and the data were not used in our regression analysis.}

\end{deluxetable*}

  \section{RESULTS}

  \label{s-results}

\begin{figure}
  \begin{center}
    \scalebox{1.0}{\rotatebox{90}{\plotone{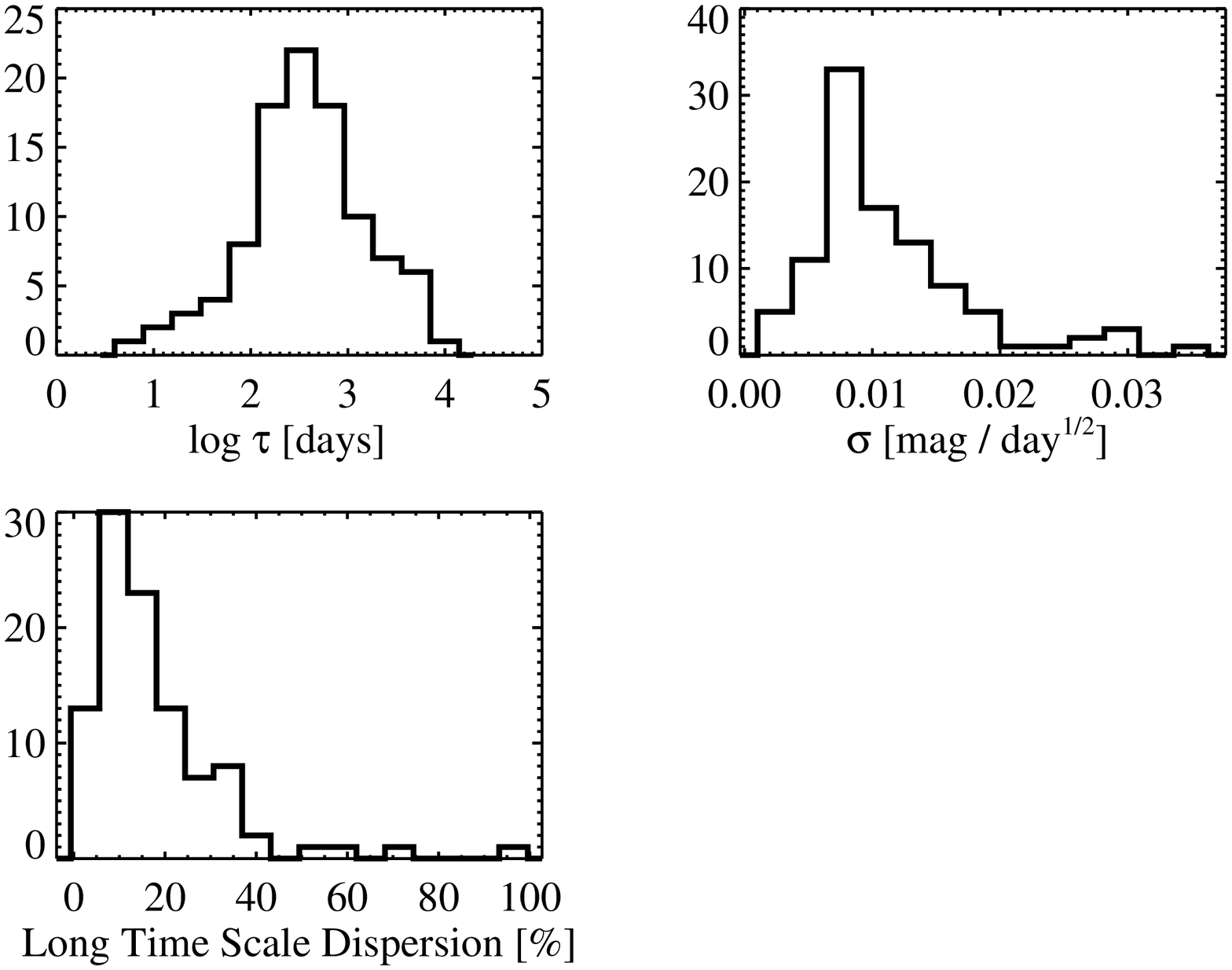}}}
    \caption{Distribution of the best fit CAR(1) process parameters
      for the 100 quasars in our sample. The characteristic time scales of
      AGN optical light curves are $10 \lesssim \tau \lesssim 10^4$
      days, the amplitudes of short time scale variations are $\sigma
      \lesssim 0.02\ {\rm mag\ day^{-1/2}}$, and the amplitudes of
      long time scale variations are $\lesssim 40\%$. The
      uncertainties on the characteristic time scales are large, and
      the true dispersion in $\tau$ is likely $\sim 0.3$ dex.
    \label{f-lchists}}
  \end{center}
\end{figure}

  The distribution of $\sigma, \tau,$ and light curve standard
  deviations for our sample are shown in Figure \ref{f-lchists}. The
  light curve standard deviation is calculated as the square root of
  the light curve variance, $s = \sigma \sqrt{\tau / 2}$. The best fit
  quasar relaxation times have a median value of 540 days and a
  dispersion of 0.64 dex, and show typical long time scale optical
  variations of $3$--$30$ per cent. However, the uncertainties on
  $\tau$ are large and make a considerable contribution to the
  observed scatter. Correcting for the contribution from the
  uncertainties implies an intrinsic dispersion in relaxation times of
  $\sim 0.3$ dex.

  In order to look for correlations of quasar variability properties
  on luminosity, redshift, black hole mass, and Eddington ratio, we
  used the linear regression method of \citet{linmix}. The method of
  \citet{linmix} takes a Bayesian approach to linear regression, and
  accurately accounts for intrinsic scatter in the regression
  relationships, as well as measurement errors in both the dependent
  and independent variables. Measurement errors can be large for both
  the estimates of $\tau$ and $M_{BH}$, and thus can have a
  significant effect on the observed correlations
  \citep[e.g.,][]{kelly07}. Therefore, it is necessary to correct for
  the measurement errors when attempting to recover any underlying
  trends.

  \subsection{Dependence of Quasar Variability on Luminosity and Redshift}

  \label{s-var_vs_lumz}

  In order to investigate whether quasar variability properties depend
  on luminosity, redshift, or both, we performed a regression
  analysis. Throughout this section, the luminosity will always be
  taken to be $\lambda L_{\lambda}$ at 5100\AA. There has been
  considerable debate over whether quasar variability is correlated
  with luminosity or redshift, and the artificial correlation between
  the two has made it difficult for previous work to uncover the true
  intrinsic correlation. However, because we perform a linear
  regresion of variability properties on both $L$ and $z$
  simultaneously, we are able to break the degeneracy between $L$ and
  $z$. This is because the multiple linear regression describes how
  variability depends on $L$ at a given $z$, and likewise for $z$ at a
  given $L$.

\begin{figure*}
  \begin{center}
    \scalebox{0.7}{\rotatebox{90}{\plotone{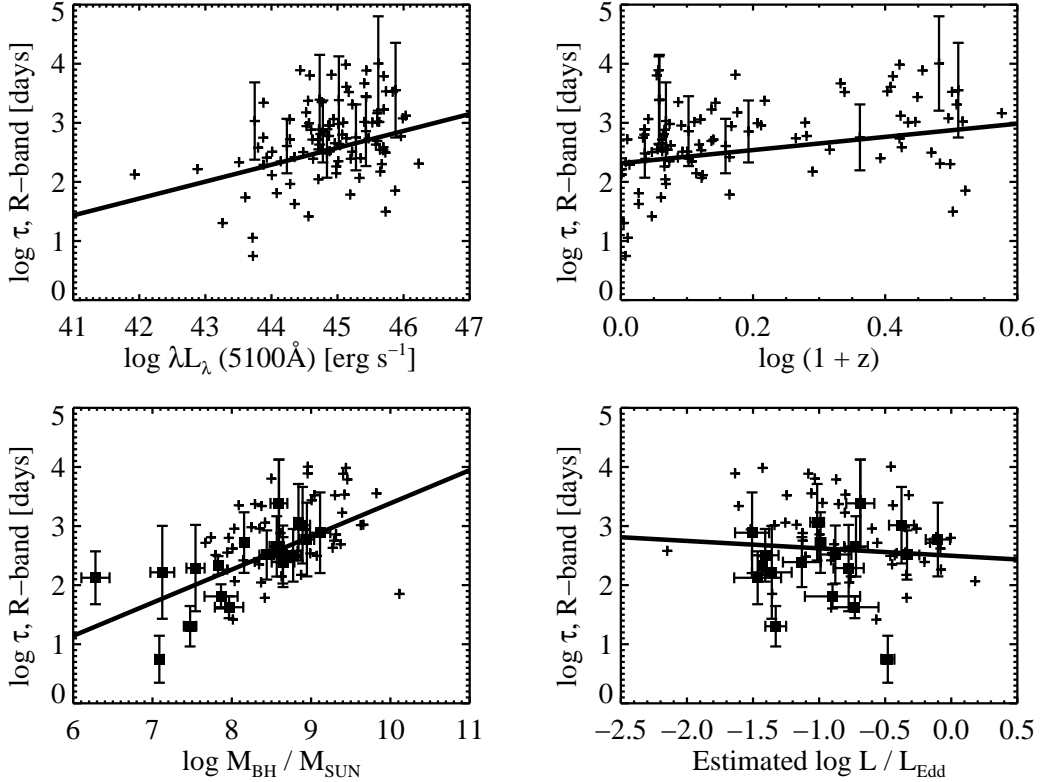}}}
    \caption{The characteristic time scale of the optical light curves
    for the AGN in our sample as a function of optical luminosity,
    redshift, black hole mass, and estimated Eddington ratio. For
    clarity, we only show error bars for a random fraction of the data
    points in the top two panels, and we only show the error bars for
    the sources with $M_{BH}$ estimated from reverberation mapping in
    the bottom two panels. The straight lines denote the best fit
    linear regression. There is a significant trend for $\tau$ to
    increase with increasing $M_{BH}$, and less significant trends
    between $\tau$ and $\lambda L_{\lambda}$ or $z$. There is no
    significant trend between $\tau$ and $L / L_{Edd}$.
    \label{f-tau}}
  \end{center}
\end{figure*}

\begin{figure*}
  \begin{center}
    \scalebox{0.7}{\rotatebox{90}{\plotone{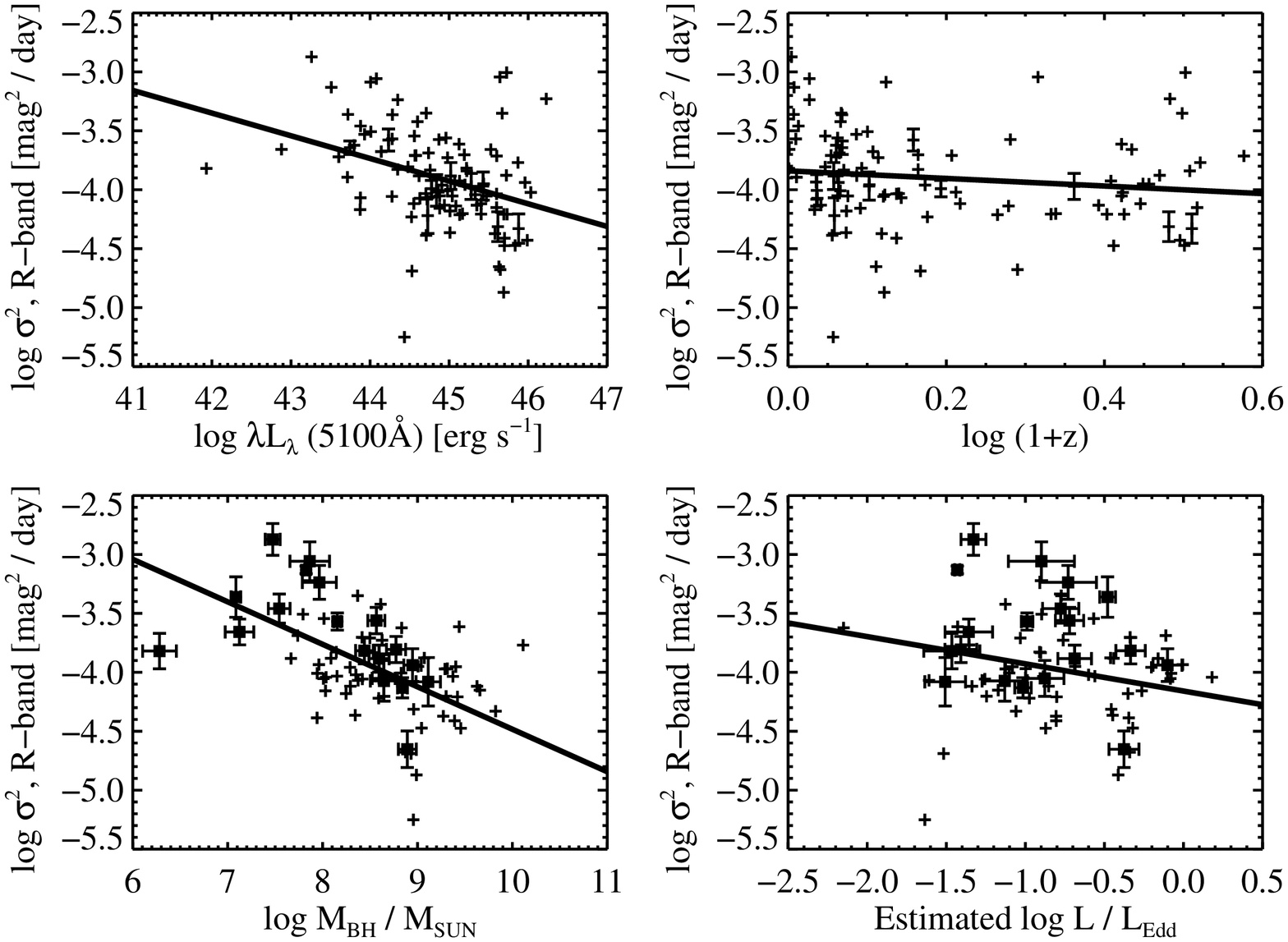}}}
    \caption{Same as Figure \ref{f-tau}, but for the variance in the
    short time scale variations, $\sigma^2$. There is a significant
    trend for $\sigma$ to decrease with increasing $M_{BH}$, and a
    similar but less significant trend between $\sigma$ and $\lambda
    L_{\lambda}$.
    \label{f-sigma}}
  \end{center}
\end{figure*}

  In Figure \ref{f-tau} we show the relaxation time $\tau$ as a
  function of luminosity and redshift, and in Figure \ref{f-sigma} we
  show $\sigma$ as a function of luminosity and redshift. The results
  of the regressions for $\tau$ are
  \begin{eqnarray}
    \log \tau & = & (-10.29 \pm 3.76) + (0.29 \pm 0.08) \log \lambda L_{\lambda}\ \ {\rm [days]}
       \label{eq-ltau} \\
    \log \tau & = &   (2.32 \pm 0.10) + (1.12 \pm 0.41) \log (1 + z) \ \ {\rm [days]}
       \label{eq-ztau} \\
    \log \tau & = & (-8.13 \pm 0.12) + (0.24 \pm 0.12) \log \lambda
    L_{\lambda} \nonumber \\
    & + & (0.34 \pm 0.58) \log (1 + z)\ \ {\rm [days]} \label{eq-lztau}, \\
  \end{eqnarray}
  and the results of the regression for $\sigma$ are
  \begin{eqnarray}
    \log \sigma^2 & = & (4.73 \pm 2.34) - (0.19 \pm 0.05) \log \lambda L_{\lambda}\ \ 
	 {\rm [R\ mag^2 / day]}
       \label{eq-lsigma} \\
    \log \sigma^2 & = & (-3.84 \pm 0.06) \nonumber \\
                  & - & (0.32 \pm 0.25) \log (1 + z) \ \ {\rm [R\ mag^2 / day]}
       \label{eq-zsigma} \\
    \log \sigma^2 & = &  (8.00 \pm 3.29) - (0.27 \pm 0.07) \log
    \lambda L_{\lambda} \nonumber \\
                  & + & (0.47 \pm 0.33) \log (1 + z)\ \ {\rm [R\ mag^2 / day]} \label{eq-lzsigma}. \\
  \end{eqnarray}
  There is a statistically significant correlation between $\tau$ and
  both $L$ and $z$, where the light curve relaxation time increases
  with increasing $L$ and $z$. In addition, there is a statistically
  significant anti-correlation between $\sigma$ and $L$, implying that
  the short time scale variance decreases with increasing
  $L$. However, the multiple regression results show that the
  luminosity trends are the dominant ones, and that there is no
  significant trend between either $\tau$ or $\sigma$ and $z$, at a
  given $L$. This therefore implies that the observed trends with
  redshift are caused by the artificial correlation between $L$ and
  $z$ resulting from selection effects.
  
  We also looked for trends of the light curve variance, $\tau
  \sigma^2 / 2$, with $L$ and $z$ and found statistically significant
  evidence for a correlation between the variance of the light curve
  and $z$. Both the Spearman and Kendall rank correlation statistic
  was significant at $3\sigma$, although there is considerable scatter
  in the correlation. This correlation is most likely a reflection of
  the well-known fact that quasar emission at shorter wavelengths is
  more variable \citep[e.g.,][]{vanden04}

  \subsection{Dependence of Quasar Characteristic Time Scale and Variability on $M_{BH}$}

  \label{s-var_vs_mbh}

  We also looked for trends in quasar variability properties with
  $M_{BH}$ and the Eddington ratio, $L / L_{Edd}$. In this work we
  assume a constant bolometric correction of $C_{bol} = 10$ to the
  luminosity at 5100\AA \citep{kaspi00}. However, we stress that a
  constant bolometric correction can introduce significant error in
  the Eddington ratio \citep[e.g.,][]{vasud07,kelly08}, and that our
  use of $C_{bol} = 10$ is only suggestive. Strictly speaking, what is
  being used in the following regressions is the ratio $\lambda
  L_{\lambda}(5100$\AA$) / M_{BH} $, and in general this will not
  equal the true Eddington ratio. In Figure \ref{f-tau} we also show
  $\tau$ as a function of $M_{BH}$ and $L / L_{Edd}$, and in Figure
  \ref{f-sigma} we also show $\sigma$ as a function of $M_{BH}$ and $L
  / L_{Edd}$.

  The results of the regressions for $\tau$ are
  \begin{eqnarray}
    \log \tau & = & (-2.29 \pm 1.17) + (0.56 \pm 0.14) \log M_{BH}\ \ {\rm [days]}
       \label{eq-mtau} \\
    \log \tau & = & (2.50 \pm 0.24) - (0.06 \pm 0.27) \log L / L_{bol} \ \ {\rm [days]}
       \label{eq-etau}.
  \end{eqnarray}
  and the results of the regression for $\sigma$ are
  \begin{eqnarray}
    \log \sigma^2 & = & (0.33 \pm 0.73) \nonumber \\
                  & - & (0.52 \pm 0.08) \log M_{BH}\ \ {\rm [R\ mag^2
                    / day]} \label{eq-msigma} \\
    \log \sigma^2 & = & (-4.33 \pm 0.19) \nonumber \\
                  & - & (0.25 \pm 0.22) \log L / L_{Edd} \ \ {\rm [R\
                    mag^2 / day]} \label{eq-esigma}.
  \end{eqnarray}
  Based on these regression results, there is a statistically
  significant correlation between $\tau$ and $M_{BH}$, where the
  relaxation time increases with increasing $M_{BH}$. In addition,
  there is significant evidence that $\sigma$ decreases with
  increasing $M_{BH}$. However, there is no evidence for a dependence
  of $\tau$ or $\sigma$ on the Eddington ratio.

  In this work we have employed the linear regression technique
  described in \citet{linmix}. However, this technique assumes
  symmetric errors in $\tau$, but in reality the error distribution is
  asymmetric. In order to assess whether this has any effects on our
  results, we modifed the technique of \citet{linmix} to handle
  asymmetric errors in $\tau$, and used our modifed code to recompute
  the dependencies involving $\tau$. However, we did not notice any
  difference in the slopes by assuming symmetric or asymmetric error
  bars. This is most likely because of the assumption of a linear
  relationship between $\log \tau$ and the other quantities. The
  asymmetric error bars for the high-mass quasars imply that
  the variability time scales for these sources are consistent with
  time scales much longer than the span of the lightcurve, which would
  lead to a steepening in the slope. However, this steepening of the
  slope would be inconsistent with the (better determined) time scales
  for the low-mass AGN, since a steeper line would predict shorter
  time scales than are observed. So, the assumption of a linear 
  relationship, in combination with the well determined time scales at
  the low-mass end, counteracts the effects of the asymmetric error
  bars at the high-mass end. However, if we
  were to assume a nonlinear form for the dependence of $\log \tau$
  on $\log M_{BH}$, such that the slope increases for the high mass
  objects, then the asymmetric error bars would likely have a
  noticeable effect, since the low-mass objects would no longer convey
  a lot of 'information' about the regression relationship at the high
  mass end.

\begin{figure*}
  \begin{center}
    \includegraphics[scale=0.33,angle=90]{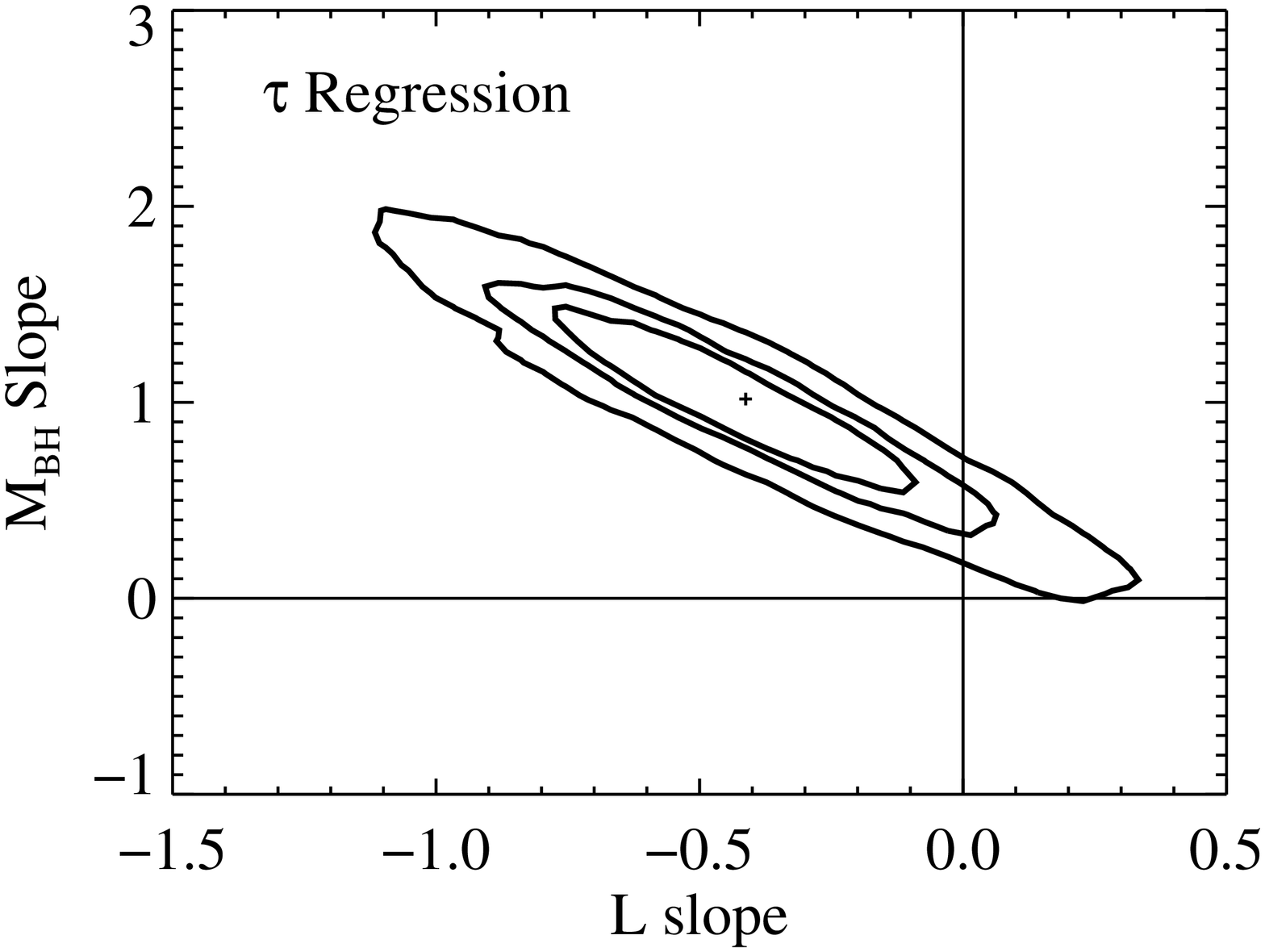}
    \includegraphics[scale=0.33,angle=90]{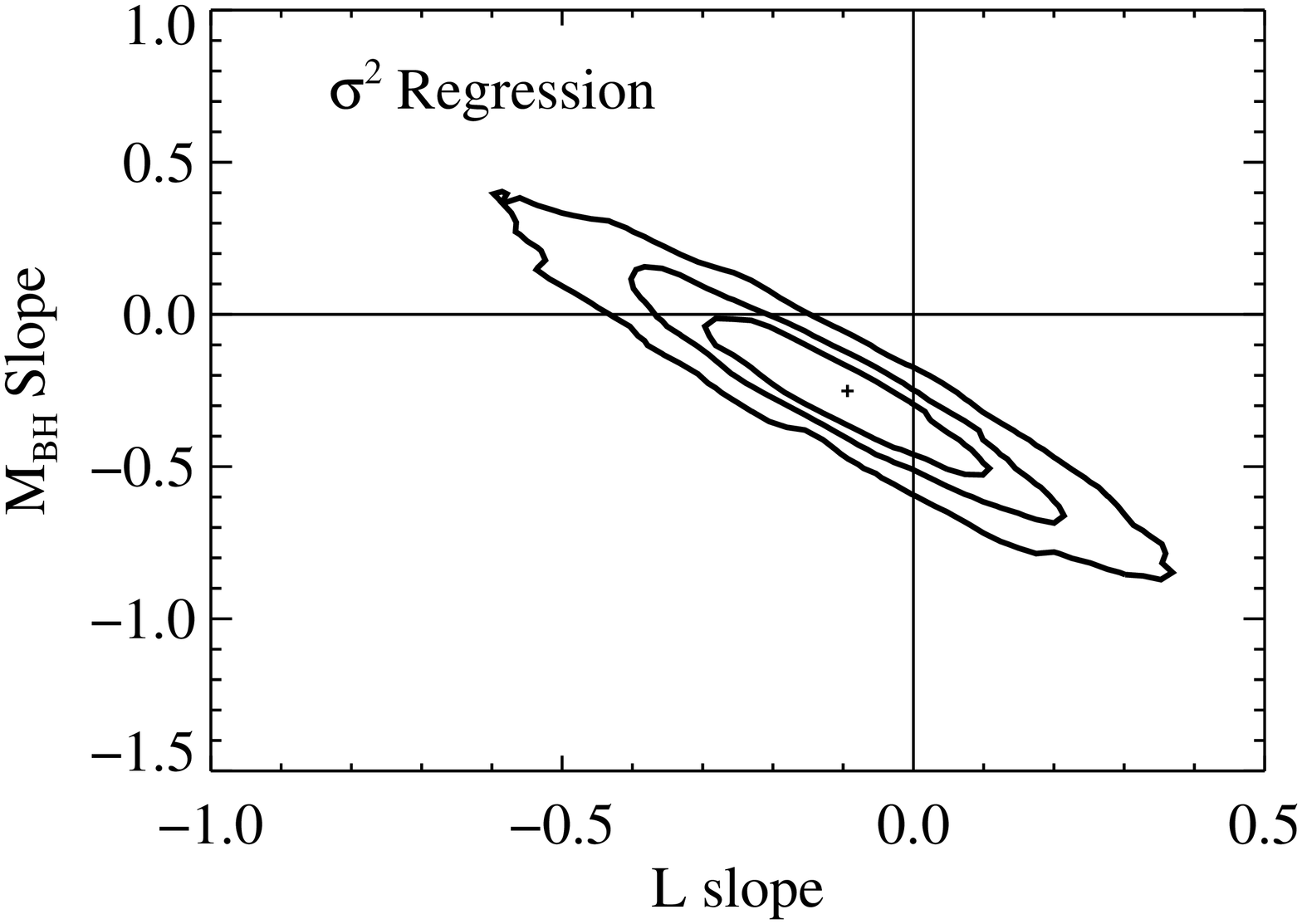}
    \caption{Posterior probability distribution for the values of the
      coefficients in a linear regression of the characteristic time
      scale of quasar optical variations, $\tau$, (left) and the
      magnitude of short time scale variations, $\sigma$, (right) as a
      function of $M_{BH}$ and $\lambda L_{\lambda}$ (see
      Eq.[\ref{eq-lmsigma}] and [\ref{eq-lmtau}]). The contours
      correspond to approximate $50\%, 75\%,$ and $95\%$ joint
      confidence regions. While there is significant evidence that
      $\tau$ depends on at least $M_{BH}$, it is unclear if there is
      an additional dependence on luminosity. In addition, while it is
      clear that $\sigma$ depends on either $M_{BH}$ or $\lambda
      L_{\lambda}$, it is unclear whether the dependency is on
      $M_{BH}$, $\lambda L_{\lambda}$, or both.
    \label{f-slopes}}
  \end{center}
\end{figure*}

  Due to the correlation between $L$ and $M_{BH}$, it is unclear
  whether the observed dependency of $\tau$ and $\sigma$ on these
  quantities is real for both $L$ and $M_{BH}$, or whether one
  correlation is simply a reflection of the other. Similar to breaking
  the $L$--$z$ degeneracy, we can investigate which correlation is the
  fundamental one, or if both are, by performing a multiple regression
  of $\tau$ and $\sigma$ on $L$ and $M_{BH}$. The results are
  \begin{eqnarray}
    \log \sigma^2 & = & (-3.83 \pm 0.17) - 
    (0.09 \pm 0.19) \log \left(\frac{\lambda L_{\lambda}}{10^{45}\ {\rm erg\ s^{-1}}}\right) \nonumber \\
     & - & (0.25 \pm 0.24) \log \left(\frac{M_{BH}}{10^8 M_{\odot}}\right)\ \ \ [{\rm R\ mag^2 / day}]
    \label{eq-lmsigma} \\
    \tau & = & (80.4^{+66.9}_{-35.8}) \left(\frac{\lambda L_{\lambda}}{10^{45}\ 
      {\rm erg\ s^{-1}}}\right)^{-0.42 \pm 0.28} \nonumber \\
         & \times & \left(\frac{M_{BH}}{10^8 M_{\odot}}\right)^{1.03 \pm 0.38}\ \ \ [{\rm days}]
    \label{eq-lmtau}
  \end{eqnarray}
  Here, we have expressed $\tau$ as a function of $L$ and $M_{BH}$
  instead of $\log \tau$ for more direct comparison with the
  characteristic time scales described by Equations
  (\ref{eq-tlc})--(\ref{eq-tth}). The joint probability distributions
  of the slopes are shown in Figure \ref{f-slopes} for both
  regressions. It is unclear whether $\sigma$ depends on solely
  $M_{BH}$, solely $L$, or both $M_{BH}$ and $L$, although the data
  favor a dependence on $M_{BH}$ over one on $L$. However, there is
  significant evidence that the relaxation time scale depends on at
  least $M_{BH}$, and possibly on $L$ as well. In fact, the dependence
  of $\tau$ on $M_{BH}$ has steepened, and the relationship described
  by Equation (\ref{eq-lmtau}) is similar to that for the orbital or
  thermal time scale, assuming a viscosity parameter of $\alpha \sim
  10^{-3}$ and location of the emitting region to be $R_S \sim 100$.

  \section{DISCUSSION}
  
  \label{s-discussion}

  \subsection{Comparison with Previous Work}

  \label{s-previous}

  Most previous work on quasar optical variability has been based on
  analysis of structure functions or power spectra, either of
  individual quasars or an ensemble of quasars. From these studies, an
  anti-correlation between variability and luminosity has often
  emerged \citep[e.g.,][]{hook94,garcia99,vanden04,wilhite08}, while
  results on a variability--redshift correlation have remained
  mixed. Our result that long-term quasar variability is uncorrelated
  with luminosity may appear to be in conflict with previous
  work. However, the evidence for a variability--luminosity
  correlation is considerably weaker in the studies that have computed
  variability measures for individual objects. Indeed, studies that
  have compared variability with luminosity for individual objects
  have noted the significant scatter in the relationship, producing a
  very weak correlation, often leading to a detection of `moderate'
  statistical significance at best. Ensemble studies, on the other
  hand, cannot investigate the scatter in the relationship and
  therefore cannot assess the strength of the correlation. Instead,
  ensemble studies simply look for an average trend in variability
  properties, and, given enough quasars, are able to detect even a
  weak trend of variability with luminosity. Furthermore, quasars
  exhibit a range in characteristic time scale and variability
  amplitude at a given luminosity or black hole mass, and it is
  unclear how this affects an ensemble structure function or power
  spectrum.

  Recently \citet{wold07} have reported a correlation between optical
  variability and $M_{BH}$ on time scales $t > 100$ days, but no
  correlation is seen on shorter time scales. In contrast, we observed
  an anti-correlation between $M_{BH}$ and short time scale
  variability, but no correlation between $M_{BH}$ and long time scale
  variability. The \citet{wold07} result is unexpected, since $M_{BH}$
  and $L$ are correlated \citep[e.g.,][]{peter04}, and previous
  studies have found that variability is anti-correlated with $L$,
  even on long time scales. The \citet{wold07} result is based on an
  ensemble structure function, and the uncertainties on the structure
  function for the high $M_{BH}$ bins are large. Furthermore, the time
  sampling of the \citet{wold07} sample for the high $M_{BH}$ bins is
  worse than for the low $M_{BH}$ bins, and windowing effects due to
  the finite length of the time series may be at work here. When
  considering variability measurement of individual sources,
  \citet{wold07} still find a positive correlation between variability
  and $M_{BH}$. However, while this correlation is statistically
  significant, it is very weak and exhibits considerable scatter. We
  performed a Kendall and Spearman rank correlation test between long
  time scale variability and the estimated black hole mass, and also
  find a marginally significant correlation, but the significance
  disappeared when we accounted for the uncertainty in the mass
  estimates. In addition, because the long time scale variability for
  a CAR(1) process is $\tau \sigma^2 / 2$, Equations (\ref{eq-mtau})
  and (\ref{eq-msigma}) imply that the long time scale variability
  depends only weakly on $M_{BH}$, if at all.

  \citet{collier01} calculated structure functions of optical light
  curves for 12 low-$z$ AGN with reverberation mapping
  data. Consistent with our work, they find a correlation between
  $M_{BH}$ and characteristic time scale, where a characteristic time
  scale was defined as the location of a break in the structure
  function. The time scales found by \citet{collier01} were consistent
  with dynamical or disk thermal time scales, although they tended to
  be somewhat shorter than those derived in this work. This difference
  may be explained by somewhat different definitions of
  `characteristic' time scale between their work and ours, and
  systematic errors in the estimated structure functions caused by
  time sampling effects.

  Two of the sources in our sample, NGC 5548 and NGC 4151, were
  studied by \citet{czerny99} and \citet{czerny03}, respectively,
  using spectral techniques. These authors fit a functional form to
  the optical power spectra similar to that implied by the CAR(1)
  model, with the exception that they allow the high-frequency slope
  to vary. In contrast, the CAR(1) model we employ fixes this slope to
  be $-2$, i.e., $P_X(f) \propto 1 / f^2$. For NGC 5548,
  \citet{czerny99} find evidence for a flattening of the optical power
  spectrum on time scales $\log \tau = 3.2 \pm 0.3$ days. This is
  longer than our observed characteristic time scale of $\log \tau =
  2.3 \pm 0.16$ at $\approx 2.5\sigma$ significance. This difference
  is only marginally significant, and may be explained by the slightly
  different parameteric forms assumed for the power spectra, the
  longer light curve used by \citet{czerny99}, the different
  photometric bands, and errors introduced by the window function and
  smoothing of the power spectra in the \citet{czerny99} analysis.
  
  For NGC 4151, \citet{czerny03} did not find any evidence for a
  flattening to white noise of the $B$-band optical power spectra over
  the time scales probed by their 90 yr light curve. However, they
  find some evidence that the power spectra flattens below time scales
  of $\tau \sim 100$ days, in agreement with the characteristic time
  scale we find of $\log \tau = 2.2 \pm 0.78$ days. Furthermore, we
  note that our analysis constrains the characteristic time scale for
  NGC 4151 to be $\lesssim 30$ yr at $99\%$ confidence. If the power
  spectra does flatten on time scales $\sim 30$ yr, it would be
  difficult to detect in the spectral analysis employed by
  \citet{czerny03}, as such a turn-over would occur in the few lowest
  frequency bins. This is especially true when one considers that the
  lowest frequency bins of the power spectra are also among the bins
  most biased by windowing effects and smoothing.

  \subsection{Connection with Accretion Physics}

  \label{s-physics}

  In this work we have found that both the characteristic time scale
  of quasar light curves, and the magnitude of the short time scale
  variations, depend on black hole mass. This, in combination with the
  evidence from reverberation mapping, strongly argues that the source
  of quasar optical variability is intrinsic to the accretion disk. In
  addition, we have found that the characteristic time scales of
  quasar light curves are similar to what would be expected for disk
  dynamical or thermal time scales, assuming a viscosity parameter of
  $\alpha \sim 10^{-3}$. Recent MHD simulations of
  radiation-dominated AGN accretion disks have found that the thermal
  time scale is shorter than that implied by the standard
  $\alpha$-prescription \citep{turner04}, making the association of
  $\tau$ with thermal time scales more consistent. Our results imply
  that on time scales shorter than $t_{orb}$ or $t_{th}$, the
  accretion disk has difficulty generating variations in optical flux
  in response to random variations of some input process, such as, for
  example, a time varying magnetic field. Instead, these short time
  scale variations get `smoothed out', creating a $1 / f^2$ power
  spectrum for frequencies higher than $\sim 1 / t_{orb}$ or $\sim 1 /
  t_{th}$.

  While the quasar characteristic time scales are consistent with both
  orbital and thermal time scales, we find it more appropriate to
  associate these time scales with $t_{th}$, as we might expect some
  sort of periodic activity in the light curves if the flux variations
  were driven by orbital motion. In addition, quasar optical emission
  is thought to be thermal emission from an optically thick accretion
  disk \citep[e.g.,][]{krolik99,frank02}, and quasars tend to be bluer
  as they brighten \citep[e.g.,][]{giveon99,trevese01}, suggesting
  that the flux variations are due to thermal variations. An
  association with orbital time scales cannot be ruled out, and
  indeed, the characteristic time scale of the turbulence driving the
  magnetic energy and kinetic energy density in the disk is $\sim
  t_{orb}$ \citep[e.g.,][]{hirose08}. However, if the radiation energy
  density in the disk is due to dissipation of magnetic and kinetic
  energy density, then it is expected to respond to fluctuations in
  the magnetic and kinetic energy densities with a characteristic time
  scale of order the thermal time.

  In order to interpret the CAR(1) process in terms of accretion disk
  physics, we rewrite Equation (\ref{eq-car1}) as
  \begin{equation}
    d\log L(t) = -\frac{1}{\tau} (\log L(t) - \mu) dt + \sigma [B(t + dt) - B(t)] 
    \label{eq-car1alt}.
  \end{equation}
  Here, $L(t)$ denotes the luminosity of the quasar at time $t$, $\mu
  = \tau b$ is the mean value of the quasar light curve, and $B(t)$
  denotes Brownian motion. In Equation (\ref{eq-car1alt}) we have used
  the fact that the derivative of Brownian motion is white noise,
  i.e., $\epsilon(t) = dB(t) = B(t+dt) - B(t)$. Brownian motion is a
  non-stationary random walk process that has a power spectrum $P(f)
  \propto 1 / f^2$, and is described by Equation (\ref{eq-car1}) in
  the limit $\tau \rightarrow \infty$. In addition, Equation
  (\ref{eq-car1alt}) implicitly assumes that the variance in the
  random variable $dB(t) = B(t + dt) - B(t)$ is $Var(dB(t)) = dt$. The
  notation $B(t)$ should not be confused with the common physics
  notation of denoting the amplitude of a magnetic field; although in
  this work we suggest that $B(t)$ may be associated with a turbulent
  magnetic field, the term $B(t)$ should be understood as referring to
  a Brownian motion.

  Writing the Equation for a CAR(1) process as Equation
  (\ref{eq-car1alt}) reveals a number of interesting properties of
  this process. First, we note that the first term on the right side
  is what keeps the time series stationary. Considering only this term
  (i.e., $\sigma = 0$), $d\log L(t) / dt$ is negative when the value
  of $L(t)$ is brighter than the mean, and $d\log L(t) / dt$ is
  positive when $L(t)$ is fainter than the mean. Therefore, the first
  term on the right side stabilizes the process by always driving
  $L(t)$ toward its mean value, while the second term generates random
  perturbations to $d\log L(t) / dt$ that cause $L(t)$ to deviate from
  its expected path. For highly accreting objects like quasars, the
  accretion disks are expected to be radiation pressure dominated in
  the regions emitting the optical and UV flux. Under the standard
  $\alpha$-prescription for the viscosity, where the viscous torque is
  assumed to be proportional to the total pressure, a radiation
  pressure dominated disk is unstable to perturbations in the heating
  rate \citep[e.g.,][]{shak76,krolik99}. The fact that the quasar
  light curves in our sample are described well by a CAR(1) process
  with relaxation times similar to disk thermal time scales rules out
  instabilities in the disk that grow as $\sim t_{th}$, consistent
  with results obtained from 3-dimensional MHD simulations
  \citep[e.g.,][]{turner04,hirose08}.

  From Equation (\ref{eq-car1alt}) it is apparent that the stochastic
  input into the differential equation, which drives the random
  variations in $L$, is itself a stochastic process. In particular, a
  random deviation in the input process, $B(t)$, over a time interval
  $dt$ causes a random perturbation to the change in $\log L(t)$
  expected over the interval $dt$, with $\sigma$ controlling how
  sensitive $d\log L(t)$ is to $dB(t)$. For example, the input process
  could be due to variations in accretion rate, or perturbations
  caused by a time-varying magnetic field. Indeed, recently some
  authors have modeled quasar X-ray variability as being driven by
  variations in a magnetic field, with the magnetic field density
  being modelled as an AR(1) process
  \citep[e.g.,][]{king04,mayer06,janiuk07}.

  \citet{hirose08} have argued for a similar interpretation of quasar
  optical variability, based on 3-dimensional MHD simulations. They
  argue that fluctuations in the magnetic field are dissipated,
  transferring energy from the magnetic field to heat in the plasma,
  creating thermal fluctuations in the accretion disk. The
  fluctuations in the heat content of the disk then create
  fluctuations in $L(t)$; however, because the disk radiation cannot
  react to changes in heat content on time scales less than the
  thermal time scale, the shorter time scale fluctuations in flux are
  smoothed out and damped. Short time scale variations in radiation
  are correlated because the radiation energy density of the disk has
  not had time to completely react to the change in heat content
  induced by the change in magnetic energy density. However, on time
  scales $t \gtrsim t_{th}$, the disk has had time to adjust to the
  heat content variations, thereby `forgetting' about the previous
  perturbations in heat content. The result is a red noise power
  spectrum on time scales $t \lesssim t_{th}$, and a white noise power
  spectrum on time scales $t \gtrsim t_{th}$.

  Within this interpretation, the parameter $\sigma$ describes the
  amplitude of variations in flux caused by variations in the magnetic
  energy density. Therefore, the anti-correlation between $\sigma$ and
  $M_{BH}$ may imply two things. First, it may imply that the
  magnitude of the stochastic input process, $B(t)$, possibly
  associated with a turbulent magnetic field, decreases with
  increasing $M_{BH}$. Or, it may imply that the sensitivity of
  changes in radiation energy density to changes in the turbulent
  magnetic field depends on $M_{BH}$, where fluctuations in the
  magnetic field create smaller fluctuations in $L(t)$ as $M_{BH}$
  increases. Alternatively, it could imply that both effects are at
  work.

  \subsection{Microvariability and Implications for the Radio-loud/Radio-quiet Dichotomy}

  Modelling quasar variability as a stochastic process provides an
  opportunity to unify both short and long time scale variability as
  the result of a single process. The source of variations in
  radio-quiet quasar optical luminosity over time scales of hours
  (so-called `microvariability' or `intranight variability') has
  remained a puzzle, although reprocessing of X-rays or a weak blazar
  component have been suggested \citep{czerny08}. In general,
  microvariability in radio-quiet quasars is not detected above the
  photometric uncertainty \citep[e.g.,][]{gupta05,carini07}; however,
  for those sources for which it is detected the standard deviation in
  the variability over the course of a night is $\sim 0.01$ mag
  \citep{gopal03,stalin04,stalin05,gupta05}. As noted in the Appendix,
  for a CAR(1) process the standard deviation of short time scale
  variations is $\approx \sigma \sqrt{\Delta t}$. As can be seen from
  figure \ref{f-lchists}, our best-fitting CAR(1) processes for quasar
  light curves predict variations of $\lesssim 0.02$ mag over 8 hours,
  consistent with what has been observed for radio-quiet
  quasars. Furthermore, it implies that the amplitude of
  microvariability should decrease with increasing black hole mass.

\begin{figure}
  \begin{center}
    \scalebox{0.8}{\rotatebox{90}{\plotone{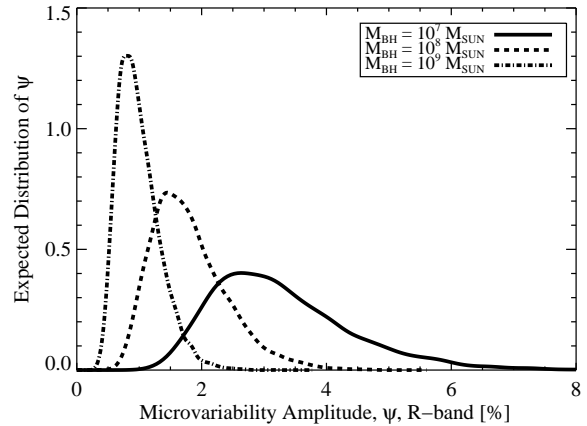}}}
    \caption{Expected probability distribution of radio-quiet quasar
      microvariability amplitude, $\psi$, predicted from the best-fit
      CAR(1) process parameters, for $M_{BH} = 10^7 M_{\odot}$ (solid
      line), $M_{BH} = 10^8 M_{\odot}$ (dashed line), and $M_{BH} =
      10^9 M_{\odot}$ (dashed-dotted line). The parameter $\psi$ is
      defined to be the maximum observed difference in flux values
      observed over a 5 hour period, in per cent. The amplitude of
      microvariability predicted from our best-fitting CAR(1)
      processes is consistent with what has been observed in
      radio-quiet quasars, implying that the same process drives both
      the long and short time scale variations in these objects.
    \label{f-psidist}}
  \end{center}
\end{figure}

  To further investigate whether our best-fitting CAR(1) processes are
  consistent with observed microvariability of radio-quiet quasars, we
  calculate the expected distribution of microvariability amplitude
  for three different values of black hole mass: $M_{BH} = 10^7,
  10^8,$ and $10^9 M_{\odot}$. We simulate light curves using the
  values of $\tau$ and $\sigma$ calculated from Equations
  (\ref{eq-mtau}) and (\ref{eq-msigma}) for each value of
  $M_{BH}$. Each simulated light curve had 40 data points, regularly
  sampled over the course of 5 hrs; these values were chosen to be
  consistent with recent observations of microvariability
  \citep[e.g.,][]{stalin05,gupta05}. For simplicity we neglect
  observational error. Following \citet{romero99} we calculate the
  amplitude of microvariability from the difference in the maximum and
  minimum observed flux values, expressed as a per cent:
  \begin{equation}
    \psi = \frac{100}{\overline{L(t)}}\left( L_{max}(t) - L_{min}(t)\right) \label{eq-psi}.
  \end{equation}
  Here, $\overline{L(t)}$ is the average value of the light curve. The
  results are shown in Figure \ref{f-psidist}. As can be seen, the
  amplitudes of microvariability expected from our best fit CAR(1)
  processes are consistent with what is observed, at least for
  radio-quiet quasars
  \citep[e.g.,][]{gopal03,stalin04,stalin05,gupta05}.

\begin{figure}
  \begin{center}
    \scalebox{0.8}{\rotatebox{90}{\plotone{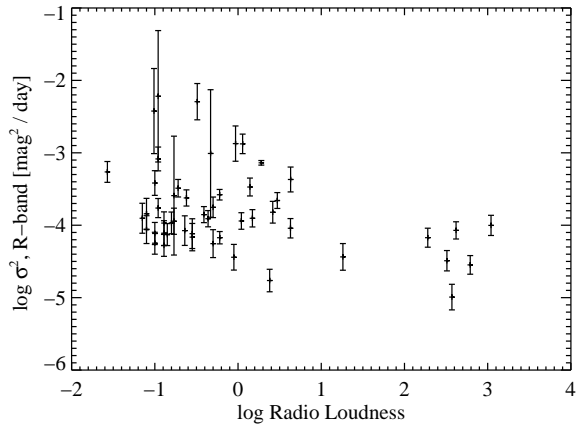}}}
    \caption{The best-fit values of $\sigma$ as a function of
      radio-loudness, $R_{6cm}$. For the CAR(1) process assumed in
      this work, short-time scale variations have a dispersion of
      $\sim \sigma \sqrt{dt}$ on time scales $\approx dt$. The
      amplitude of microvariability decreases with increasing
      radio-loudness, opposite the trend found from direct
      observations of microvariability in radio-quiet and radio-loud
      sources. This, along with the fact that our study only probes
      time scales $\gtrsim 2$ days, implies that radio-loud objects
      have an additional variability component that operates on
      intranight time scales.
    \label{f-rloud_vs_sigma}}
  \end{center}
\end{figure}

  Radio-loud quasars, on the other hand, are known to exhibit stronger
  microvariability, and the intranight fluctuations are often above
  the detection threshold \citep[e.g.,][]{gupta05}. If the physical
  mechanism for microvariability is the same in both radio-quiet
  quasars and radio-loud quasars, then the observational result that
  radio-loud quasars exhibit stronger microvariability implies a
  correlation between $\sigma$ and radio-loudness. To test this, we
  compare the values of the radio-loudness, $R_{6cm}$, for the
  \citet{giveon99} and AGN Watch samples. The radio-loudness
  parameter, $R_{6cm}$, is defined in the standard way to be the ratio
  of flux density at 6cm to the flux density at 4400\AA; the
  traditional division between radio-quiet and radio-loud sources
  occurs at $R_{6cm} = 10$ \citep{kell89}. Figure
  \ref{f-rloud_vs_sigma} shows the best-fit values of $\sigma^2$,
  which give the optical variance on time scales of $\approx 1$ day,
  as a function of radio-loudness. As can be seen, $\sigma$ is
  anti-correlated with radio-loudness, showing that our best-fit
  CAR(1) process parameters predict microvariability should actually
  be reduced in radio-loud quasars. This is inconsistent with the
  observational result that microvariability is stronger in radio-loud
  quasars, and implies an additional variability mechanism in
  radio-loud objects on time scales shorter than those probed by our
  study.

  Our analysis suggests that radio-loud quasars have an additional
  variability mechanism that operates on times scales $\lesssim 1$
  day. Because the time sampling for the lightcurves analyzed in this
  sample is $\gtrsim 2$ days, we are only able probe variations on
  time scales $2{\rm\ days} \lesssim \delta t \lesssim 7{\rm\ yrs}$
  when fitting the CAR(1) process. Therefore variations on time scales
  $\lesssim 1$ day are not used in estimating $\sigma$. The fact that
  the magnitude of microvariability predicted by the best-fit CAR(1)
  processes for radio-quiet quasars is consistent with direct
  observations of microvariability implies that the microvariability
  in radio-quiet quasars is caused by the same process that drives the
  longer time-scale variations. In this work we have suggested that
  this process, which is well-described as a CAR(1) stochastic
  process, is the result of a turbulent magnetic field driving thermal
  fluctuations in the accretion disk. If this process is also the
  source of micro-variability in radio-loud objects, then the
  anti-correlation between $\sigma$ and radio-loudness would imply
  that microvariability should be reduced in radio-loud
  quasars. However, because the opposite is true, this implies that
  radio-loud objects have an additional source of variability on time
  scales $\lesssim 1$ day, but does not operate on the time scales
  probed by are study, i.e., $2{\rm\ days} \lesssim \delta t \lesssim
  7{\rm\ yrs}$. This additional source of optical variability in
  radio-loud objects is likely due to the presence of a relativistic
  jet, as has often been suggested.

  Because the magnitude of microvariability predicted by our best-fit
  CAR(1) process is consistent with observations of microvariability
  in radio-quiet quasars, it is not necessary to invoke an additional
  physical mechanism to explain the short time scale variations in
  these objects. Instead, the CAR(1) process is able to explain both
  short and long time scale variations as being driven by the same
  process, possibly thermal fluctuations being driven by a turbulent
  magnetic field. \citet{czerny08} investigated three different models
  for microvariability in radio-quiet quasars and concluded that the
  most promising source of the microvariability was a weak blazar-like
  jet. Furthermore, \citet{czerny08} concluded that luminosity
  fluctuations driven by the magneto-rotational instability are not
  strong enough to create microvariability. Our empirical results are
  inconsistent with the conclusions of \citet{czerny08} in that we do
  not find any evidence for an additional process that drives the
  microvariability. This is not to say that our results imply that a
  weak jet does not exist in radio-quiet quasars, but rather that if a
  weak jet does exist then it's variability amplitude is small
  compared to that caused by the process that also drives the longer time
  scale variations, such as a turbulent magnetic field. 

  It is unclear how sensitive the conclusions of \citet{czerny08} are
  to the specifics of their model, and the investigation of
  \citet{czerny08} does not necessarily rule a turbulent magnetic
  field as being the source of microvariability in radio-quiet
  quasars. Of particular interest is the manner in which the time
  dependence of the magnetic field is modeled. \citet{czerny08}, as
  well as \citet{king04}, \citet{mayer06}, and \citet{janiuk07}, have 
  modeled the time series of the magnetic field as an AR(1) process:
  \begin{equation}
    u_{i+1} = \alpha_{AR} u_i + \epsilon_i, \label{eq-bfield}
  \end{equation}
  where the amplitude of the local magnetic field at the $i^{\rm th}$
  time step is proportional to $u_{i}$, and $\epsilon_i$ is a random
  variable uniformly distributed between $\pm
  \sqrt{3}$. Because the characteristic time scale of the magnetic
  field development is $\tau_B \sim k_d \tau_{orb}$, \citep{czerny08} used a
  time step equal to the orbital time. However, Equation
  (\ref{eq-torb}), in combination with Figure 10 of \citep{czerny08},
  implies that the orbital time scale should be $\sim 100$--$1000$
  days, depending on the mass of the black hole. This results in a
  time sampling of the magnetic field that is much longer than the
  time scales over which microvariability are measured, therefore
  artificially suppressing the amplitude of microvariability in the
  simulated light curves. Moreover, the amplitude of microvariability
  can also be increased by increasing the variance of the random
  variable $\epsilon_i$.

  Following \citet{king04}, \citet{czerny08} assumed a
  value of $\alpha_{AR} = 0.5$. As described in the appendix, a value
  of $\alpha_{AR} = 0.5$ corresponds to a characteristic time scale
  for the equivalent continuous process of $\tau = 1.44 \tau_{orb}$,
  i.e., a value of $k_d = 1.44$. A more accurate simulated light curve
  can be obtained by using a sampling interval in Equation
  (\ref{eq-bfield}) that is small compared to that probed by
  microvariability, such as $\Delta t \sim$ minutes. Then, the value
  of $\alpha$ can be related to the characteristic time scale of the
  magnetic field development as $\alpha_{AR} = \exp \{ -\Delta t / (k_d
  \tau_{orb}) \}$. In this case, the only free parameters are $k_d$ and
  the variance of $\epsilon_i$. Further improvement can be obtained by
  employing 3-dimensional MHD simulations to test whether a turbulent
  magnetic field can create microvariability with fluctuations $\sim
  2\%$.

  \section{SUMMARY}

  \label{s-summary}

  In this work we have modeled quasar light curves as a type of
  stochastic process called a first-order continuous autoregressive
  process. This statistical model has three free parameters: the
  characteristic time scale for the process to `forget' about itself,
  $\tau$, the magnitude of the small time scale variations, $\sigma$,
  and the mean of the time series, $\mu$. We used this model to fit
  100 quasar light curves at $z < 2.8$, including 70 quasars with
  black hole mass estimates. Our conclusions are summarized as
  follows:
  \begin{itemize}
    \item 
      Quasar optical light curves are often well described by a
      continuous autoregressive process (CAR(1)). For the quasars in
      our sample, the amplitude of variations on time scales long
      compare to their characteristic time scales (i.e., $sim$
      decades) are typically $\sim 3$--$30\%$, consistent
      with previous work. The characteristic time scales of the quasar
      light curves vary between $\sim 10$ days and $\sim 10$ yrs and
      are consistent with accretion disk orbital or thermal time
      scales, assuming a viscosity parameter of $\alpha \sim
      10^{-3}$ and location of the emitting region to be $R_S \sim
      100$. In addition, the short time scale variations are 
      $\lesssim 0.03\ {\rm mag\ day^{-1/2}}$.
    \item
      The characteristic time scales of quasar optical light curves
      are correlated with $M_{BH}$ and luminosity, while the magnitude
      of the short time scale variations are anti-correlated with
      $M_{BH}$ and luminosity. We did not find any evidence for an
      additional redshift correlation. A multiple regression analysis
      suggested that the primary correlation is with $M_{BH}$. At a
      given luminosity, the characteristic time scales depend on
      $M_{BH}$ as
      \begin{eqnarray}
	\tau & = & (80.4^{+66.9}_{-35.8}) \left(\frac{\lambda
	  L_{\lambda}}{10^{45}\ {\rm erg\ s^{-1}}}\right)^{-0.42 \pm
        0.28} \nonumber \\
	& \times & \left(\frac{M_{BH}}{10^8 M_{\odot}}\right)^{1.03 \pm 0.38}\ \ \
	     [{\rm days}] \nonumber
      \end{eqnarray}
      where the errors are quoted at $68\%$ confidence ($1\sigma$).
    \item
      Within the CAR(1) process model of quasar light curves, the
      random perturbations to $d\log L(t)$ are caused by a stochastic
      process. This stochastic input process may be a turbulent
      magnetic field that creates fluctuations in the accretion disk
      radiation energy density. Variations in the input process over
      an interval $dt$ create variations in $L(t)$ which are smoothed
      out on time scales shorter than an orbital or thermal time
      scale.
    \item
      The fact that radio-quiet quasar optical light curves can be
      well fit by a CAR(1) model suggests that it is not necessary to
      invoke an additional physical mechanism to describe short time
      scale variations in these objects. Instead, within the CAR(1)
      model, both short and long time scale variations are driven by
      an underlying stochastic process that causes random
      perturbations to $d\log L(t) / dt$. This is supported by the
      fact that the intranight variations predicted by our best fit
      CAR(1) processes are $\sim 2\%$ in the $R$-band, consistent with
      observations of intranight variability. Furthermore, our
      best-fit CAR(1) parameters predict an anti-correlation between
      radio-loudness and microvariability, inconsistent with what has
      been observed. This suggests that radio-loud quasars have an
      additional optical variability mechanism that operates on time
      scales shorter than those probed by our study, $\lesssim 2$
      days.
  \end{itemize}

  Before concluding, we stress that the CAR(1) model is a statistical
  model and not a physical model. Quasar light curves are stochastic
  in nature \citep[e.g.,][]{czerny97}, and the dependence of
  luminosity on the time-varying properties of the disk is complex. In
  this sense, the randomness in the stochastic model is not due to
  that fact that the physical processes themselves are not
  deterministic, but rather is a reflection of our lack of knowledge
  of the complex physical processes that generate variations in
  flux. While a physical model is needed in order to interpret the
  stochastic model in terms of accretion disk physics, and thus lead
  to a proper understanding of quasar light curves, the stochastic
  model is sufficient for modeling the data, given our current
  knowledge. Furthermore, much of the mathematical formalism of
  accretion physics is in the language of differential equations,
  suggesting that stochastic differential equations are a natural
  choice for modeling quasar light curves.

  The field of stochastic processes is a rich field with
  well-developed methodology, predominantly because of its importance
  in financial and economic modeling. We have utilized the CAR(1)
  model because of its simplicity, and because it allows us to perform
  statistical inference without having our results biased by the
  irregular sampling, measurement errors, and finite span of the time
  series. However, the CAR(1) model is the simplest of stationary
  continuous autoregressive processes, and additional flexibility may
  be achieved through the addition of higher order derivatives to
  Equation (\ref{eq-car1}). This provides a rich and flexible method
  of modeling the power spectra of quasar light curves without
  suffering from the windowing effects that can bias traditional
  Fourier and structure function techniques. For example,
  quasi-periodic oscillations can be modeled through the addition of
  second order derivatives to Equation (\ref{eq-car1}), as has been
  done in the analysis of the frequency of sun spot numbers
  \citep{phadke74}. In addition, Equation (\ref{eq-car1}) can be
  generalized to a vector form, allowing the simultaneous modeling of
  quasar light curves across multiple observing wavelengths, and thus
  introducing additional constraints on physical models of quasar
  variability. Both the use of higher order terms and multiwavelength
  modelling of quasar light curves will be the subject of future
  research.

  A computer routine for fitting the CAR(1) model to a time series is
  available on request from B. Kelly.

  \acknowledgements

  We would like to thank Tim Axelrod for supplying the \emph{MACHO}
  light curves, and Marla Geha for providing optical spectra for a
  number of the \emph{MACHO} sources and providing helpful comments on
  an early draft of this paper. We would also like to thank the
  referee, Bo\.{z}ena Czerny, for a careful reading of the manuscript
  and comments that led to its improvement. BK acknowledges support from 
  NASA through Hubble Fellowship grant \#HF-01220.01 awarded by the
  Space Telescope Science Institute, which is operated by the
  Association of Universities for Research in Astronomy, Inc., for
  NASA, under contract NAS 5-26555. AS acknowledges support from NASA
  contract NAS 8-39073.

  \appendix

  \section{Description of Autoregressive Processes}

  \label{s-car1_intro}

  The first order autoregressive process (AR(1)) process is a
  well-studied stochastic process \citep[e.g., see][and references
    therein]{scargle81}, generated according to
  \begin{equation}
    x_i = \alpha_{AR} x_{i-1} + \epsilon_i,
    \label{eq-ar1}
  \end{equation}
  where $\epsilon_i$ is a normally-distributed random variable with
  zero mean and variance $\sigma_{AR}^2$, and the data $x_i$ are
  observed at regular time intervals. The parameters of the AR(1)
  model are $\alpha_{AR}$ and $\sigma_{AR}^2$, and $\alpha_{AR}$ is
  usually constrained as $|\alpha_{AR}| < 1$ in order to ensure
  stationarity; a time series is said to be stationary when its mean
  and covariance do not vary with time\footnote[2]{Technically, this
    is referred to as 'weakly stationary', while stationarity in the
    strictest sense implies that the probability distribution of the
    time series does not change with time. However, because we are
    only concerned with Gaussian processes in this work, the two
    definitions are equivalent}. The case $\alpha_{AR}$ = 1
  corresponds to a random walk. For quasar light curves, the $n$
  observed data points $x_1, \ldots, x_n$ correspond to the observed
  fluxes at times $t_1, \ldots, t_n$, for $t_i = t_1 + (i-1) \Delta
  t$.

  The discrete AR(1) process is only defined for regularly sampled
  time series. However, astronomical time series are rarely regularly
  sampled, and often large gaps in time can exist. Furthermore, the
  AR(1) process is a discrete process, but the underlying physical
  process that gives rise to the observed flux is continuous. Because
  of these two considerations, we instead model the quasar light
  curves as a first order continuous autoregressive (CAR(1))
  process. As noted in \S~\ref{s-car1}, the CAR(1) process is
  described by the following stochastic differential equation
  \citep[e.g.,][]{brockwell02}:
  \begin{equation}
    dX(t) = -\frac{1}{\tau} X(t) dt + \sigma \sqrt{dt} \epsilon(t) + b\ dt,\ \ \ \tau, \sigma, t > 0,
    \label{eq-acar1}
  \end{equation}
  where, $\tau$ is called the `relaxation time' of the process $X(t)$,
  and $\epsilon(t)$ is a white noise process with zero mean and
  variance equal to one. Throughout this work we have assumed that the
  white noise process is also Gaussian. In the physics literature,
  Equation (\ref{eq-acar1}) is often referred to as an
  Ornstein-Uhlenbeck (O-U) process, and plays a central role in the
  mathemematics of Brownian motion; see \citet{gill96} for a review of
  the O-U process.

  The solution to Equation (\ref{eq-acar1}) is
  \begin{equation}
    X(t) = e^{-t / \tau} X(0) + b\tau (1 - e^{-t / \tau}) + 
    \sigma \int_{0}^{t} e^{-(t - s) / \tau} dB(s).
    \label{eq-car1sol}
  \end{equation}
  Here, $X(0)$ is a random variable describing the initial value of
  the time series, $dB(s)$ is a temporally uncorrelated normally
  distributed random variable with zero mean and variance
  $dt$. Strictly speaking, $dB(t)$ is an interval of Brownian motion,
  and the integral on the right side represents the stochastic
  component of the time series. From Equation (\ref{eq-car1sol}) it
  can be seen that if $\sigma = 0$, i.e., if there is no stochastic
  component, and if $X(0)$ represents a random perturbation, $X(t)$
  relaxes to it's mean value with an $e$-folding time scale $\tau$;
  hence the identification of $\tau$ as the relaxation time. If
  $\sigma > 0$, then the path that $X(t)$ takes will vary randomly
  about the expected exponential relaxation.

  The expected value of $X(t)$ given $X(s)$ for $s < t$ is
  \begin{equation}
    E(X(t)|X(s)) = e^{-\Delta t / \tau} X(s) + b\tau (1 - e^{-\Delta t / \tau})
    \label{eq-cexpect}
  \end{equation}
  and the variance in $X(t)$ given $X(s)$ is
  \begin{equation}
    Var(X(t)|X(s)) = \frac{\tau \sigma^2}{2} \left[1 - e^{-2\Delta t / \tau} \right]
    \label{eq-cvar}
  \end{equation}
  where $\Delta t = t - s$. If $2\Delta t / \tau \ll 1$, then Equation
  (\ref{eq-cvar}) implies that the variance on time scales much
  shorter than the relaxation time is $\approx \sigma^2 \Delta
  t$. Therefore, $\sigma^2$ can be interpreted as representing the
  variance in the light curve on short time scales. In addition, one
  can show that when the time sampling is regular with $\Delta t = 1$,
  then the CAR(1) process reduces to an AR(1) process with
  $\alpha_{AR} = e^{-1 / \tau}$ and $\sigma_{AR}^2 = \tau \sigma^2(1 -
  e^{-2 / \tau}) / 2$.

  In astronomical time series analysis it is common to interpret a
  light curve in terms of its autocorrelation function and power
  spectrum. The autocovariance function at time $t'$ is defined to be
  the expected value of the product of $X(t)$ and $X(t + t')$, and the
  autocorrelation function is calculated by dividing the
  autocovariance function by the variance of the time series. The
  autocorrelation function of the CAR(1) process is
  \begin{equation}
    ACF(t') = e^{-t' / \tau}.
    \label{eq-acf}
  \end{equation}
  Equation (\ref{eq-acf}) states that the correlations in CAR(1) light
  curve fall off exponentially with lag $t'$, with an $e$-folding time
  equal to the relaxation time, $\tau$. Following \citet{gill96}, the
  power spectrum of a process is computed from the autocovariance
  function $r_X(t')$ as
  \begin{equation}
    P_{X}(f) = 4 \int_{0}^{\infty} r_X(t') \cos(2 \pi f t')\ dt', \ \ \ f \ge 0.
    \label{eq-pspec_def}
  \end{equation}
  For a CAR(1) process, $r_X(t') = (\tau \sigma^2 / 2) e^{-t' /
    \tau}$, from which it follows that the power spectrum of a CAR(1)
  process is
  \begin{equation}
    P_{X}(f) = \frac{2 \sigma^2 \tau^2}{1 + (2 \pi \tau f)^2}.
    \label{eq-apspec}
  \end{equation}
  The power spectrum of the CAR(1) process falls off as $1 / f^2$ on
  time scales short compared to the relaxation time, and flattens to
  white noise at time scales long compared to the relaxation time.


\begin{thebibliography}{}

  \bibitem[Alcock et al.(1997)]{alcock97} Alcock, C., et al.\ 1997,
    \apj, 486, 697
  \bibitem[Alcock et al.(1999)]{alcock99} Alcock, C., et al.\ 1999,
    \pasp, 111, 1539
  \bibitem[Aretxaga et al.(1997)]{aretxaga97} Aretxaga, I., Cid
    Fernandes, R., \& Terlevich, R.~J.\ 1997, \mnras, 286, 271
  \bibitem[Balbus \& Hawley(1991)]{balbus91} Balbus, S.~A., \& Hawley,
    J.~F.\ 1991, \apj, 376, 214
  \bibitem[Balbus \& Hawley(1998)]{balbus98} Balbus, S.~A., \& Hawley,
    J.~F.\ 1998, Reviews of Modern Physics, 70, 1
  \bibitem[Brockwell \& Davis(2002)]{brockwell02} Brockwell, P.~.J.,
    \& Davis, R.~A.\ 2002 Introduction to Time Series and Forecasting
    (2nd Ed.; New York, NY: Springer)
  \bibitem[Carini et al.(2007)]{carini07} Carini, M.~T., Noble, J.~C.,
    Taylor, R., \& Culler, R.\ 2007, \aj, 133, 303
  \bibitem[Carone et al.(1996)]{carone96} Carone, T.~E., et al.\ 1996,
    \apj, 471, 737
  \bibitem[Cid Fernandes et al.(1996)]{cid96} Cid Fernandes, R.~J.,
    Aretxaga, I., \& Terlevich, R.\ 1996, \mnras, 282, 1191
  \bibitem[Cid Fernandes et al.(2000)]{cid00} Cid Fernandes, R.,
    Sodr{\'e}, L., Jr., \& Vieira da Silva, L., Jr.\ 2000, \apj, 544,
    123
  \bibitem[Collier et al.(1998)]{collier98} Collier, S.~J., et
    al.\ 1998, \apj, 500, 162
  \bibitem[Collier \& Peterson(2001)]{collier01} Collier, S., \&
    Peterson, B.~M.\ 2001, \apj, 555, 775
  \bibitem[Cristiani et al.(1990)]{crist90} Cristiani, S., Vio, R., \&
    Andreani, P.\ 1990, \aj, 100, 56
  \bibitem[Cristiani et al.(1996)]{crist96} Cristiani, S., Trentini,
    S., La Franca, F., Aretxaga, I., Andreani, P., Vio, R., \& Gemmo,
    A.\ 1996, \aap, 306, 395
  \bibitem[Cutri et al.(1985)]{cutri85} Cutri, R.~M., Wisniewski,
    W.~Z., Rieke, G.~H., \& Lebofsky, M.~J.\ 1985, \apj, 296, 423
  \bibitem[Czerny et al.(2003)]{czerny03} Czerny, B., Doroshenko,
    V.~T., Niko{\l}ajuk, M., Schwarzenberg-Czerny, A., Loska, Z., \&
    Madejski, G.\ 2003, \mnras, 342, 1222
  \bibitem[Czerny \& Lehto(1997)]{czerny97} Czerny, B., \& Lehto,
    H.~J.\ 1997, \mnras, 285, 365
  \bibitem[Czerny et al.(1999)]{czerny99} Czerny, B.,
    Schwarzenberg-Czerny, A., \& Loska, Z.\ 1999, \mnras, 303, 148
  \bibitem[Czerny et al.(2008)]{czerny08} Czerny, B., Siemiginowska,
    A., Janiuk, A., \& Gupta, A.~C.\ 2008, \mnras, 386, 1557
  \bibitem[de Vries et al.(2005)]{devries05} de Vries,
    W.~H., Becker, R.~H., White, R.~L., \& Loomis, C.\ 2005, \aj,
    129, 615
  \bibitem[di Clemente et al.(1996)]{diclemente96} di Clemente, A.,
    Giallongo, E., Natali, G., Trevese, D., \& Vagnetti, F.\ 1996,
    \apj, 463, 466
  \bibitem[Frank, King, \& Raine(2002)]{frank02} Frank, J., King, A.,
    \& Raine, D.\ 2002, Accretion Power in Astrophysics (3rd
    Edition;Cambridge, UK:Cambridge Univ. Press)
  \bibitem[Garcia et al.(1999)]{garcia99} Garcia, A., Sodr{\'e}, L.,
    Jablonski, F.~J., \& Terlevich, R.~J.\ 1999, \mnras, 309, 803
  \bibitem[Geha et al.(2003)]{geha03} Geha, M., et al.\ 2003, \aj,
    125, 1
  \bibitem[Gillespie(1995)]{gill96} Gillespie, D.~T.\ 1996,
    Am.~J.~Phys., 64, 225
  \bibitem[Giveon et al.(1999)]{giveon99} Giveon, U., Maoz, D.,
    Kaspi, S., Netzer, H., \& Smith, P.~S.\ 1999, \mnras, 306, 637
  \bibitem[Gopal-Krishna et al.(2003)]{gopal03} Gopal-Krishna, Stalin,
    C.~S., Sagar, R., \& Wiita, P.~J.\ 2003, \apjl, 586, L25
  \bibitem[Gupta \& Joshi(2005)]{gupta05} Gupta, A.~C., \& Joshi,
    U.~C.\ 2005, \aap, 440, 855
  \bibitem[Hawkins(2000)]{hawkins00} Hawkins, M.~R.~S.\ 2000, \aaps,
    143, 465
  \bibitem[Hawkins(2007)]{hawkins07} Hawkins, M.~R.~S.\ 2007, \aap,
    462, 581
  \bibitem[Helfand et al.(2001)]{helfand01} Helfand, D.~J., Stone,
    R.~P.~S., Willman, B., White, R.~L., Becker, R.~H., Price, T.,
    Gregg, M.~D., \& McMahon, R.~G.\ 2001, \aj, 121, 1872
  \bibitem[Hirose et al.(2008)]{hirose08} Hirose, S., Krolik, J.~H.,
    \& Blaes, O.\ 2008, in press at \apj (arXiv:0809.1708)
  \bibitem[Hook et al.(1994)]{hook94} Hook, I.~M., McMahon, R.~G.,
    Boyle, B.~J., \& Irwin, M.~J.\ 1994, \mnras, 268, 305
  \bibitem[Janiuk et al.(2000)]{janiuk00} Janiuk, A., Czerny, B., 
    \& Siemiginowska, A.\ 2000, \apjl, 542, L33
  \bibitem[Janiuk et al.(2002)]{janiuk02} Janiuk, A., Czerny, B., 
    \& Siemiginowska, A.\ 2002, \apj, 576, 908 
  \bibitem[Janiuk \& Czerny(2007)]{janiuk07} Janiuk, A., \&
    Czerny, B.\ 2007, \aap, 466, 793
  \bibitem[Kaspi et al.(1996)]{kaspi96} Kaspi, S., et al.\ 1996, \apj,
    470, 336
  \bibitem[Kaspi et al.(2000)]{kaspi00} Kaspi, S., Smith, P.~S., 
    Netzer, H., Maoz, D., Jannuzi, B.~T., \& Giveon, U.\ 2000, \apj, 533, 631 
  \bibitem[Kawaguchi et al.(1998)]{kawa98} Kawaguchi, T., 
    Mineshige, S., Umemura, M., \& Turner, E.~L.\ 1998, \apj, 504, 671 
  \bibitem[Kellermann et al.(1989)]{kell89} Kellermann, K.~I., Sramek,
    R., Schmidt, M., Shaffer, D.~B., \& Green, R.\ 1989, \aj, 98, 1195
  \bibitem[Kelly(2007)]{linmix} Kelly, B.~C.\ 2007, \apj, 665, 1489
  \bibitem[Kelly \& Bechtold(2007)]{kelly07} Kelly, B.~C., \&
    Bechtold, J.\ 2007, \apjs, 168, 1
  \bibitem[Kelly et al.(2008)]{kelly08} Kelly, B.~C., Bechtold, J.,
    Trump, J.~R., Vestergaard, M., \& Siemiginowska, A.\ 2008, \apjs,
    176, 355
  \bibitem[King et al.(2004)]{king04} King, A.~R., Pringle, J.~E.,
    West, R.~G., \& Livio, M.\ 2004, \mnras, 348, 111
  \bibitem[Krolik(1999)]{krolik99} Krolik, J.~H.\ 1999, Active
    Galactic Nuclei: From the Central Engine to the Galactic Environment
    (Princeton, NJ:Princeton Univ. Press)
  \bibitem[Lightman \& Eardley(1974)]{lightman74} Lightman, A.~P., \&
    Eardley, D.~M.\ 1974, \apjl, 187, L1
  \bibitem[Lub \& de Ruiter(1992)]{lub92} Lub, J., \& de Ruiter,
    H.~R.\ 1992, \aap, 256, 33 
  \bibitem[Lyubarskii(1997)]{lyub97} Lyubarskii, Y.~E.\ 1997, \mnras,
    292, 679
  \bibitem[Manmoto et al.(1996)]{manmoto96} Manmoto, T., Takeuchi, M.,
    Mineshige, S., Matsumoto, R., \& Negoro, H.\ 1996, \apjl, 464,
    L135
  \bibitem[Mayer \& Pringle(2006)]{mayer06} Mayer, M., \& Pringle,
    J.~E.\ 2006, \mnras, 368, 379
  \bibitem[Merloni(2003)]{merloni03} Merloni, A.\ 2003, \mnras, 341,
    1051
  \bibitem[Merloni \& Fabian(2002)]{merloni02} Merloni, A., \& Fabian,
    A.~C.\ 2002, \mnras, 332, 165
  \bibitem[Nayakshin et al.(2000)]{nayak00} Nayakshin, S., 
    Rappaport, S., \& Melia, F.\ 2000, \apj, 535, 798 
  \bibitem[Pessah et al.(2008)]{pessah08} Pessah, M.~E., Chan, C.-K., \&
    Psaltis, D.\ 2008, \mnras, 383, 683
  \bibitem[Peterson et al.(2000)]{peter00} Peterson, B.~M., et
    al.\ 2000, \apj, 542, 161
  \bibitem[Peterson et al.(2002)]{peter02} Peterson, B.~M., et
    al.\ 2002, \apj, 581, 197
  \bibitem[Peterson et al.(2004)]{peter04} Peterson, B.~M., et al.\
    2004, \apj, 613, 682
  \bibitem[Phadke \& Wu(1974)]{phadke74} Phadke, M.~S., \& Wu, S.~M.\
    1974, J.~Amer.~Statist.~Assoc., 69, 325
  \bibitem[Richards et al.(2001)]{rich01} Richards, G.~T., et al.\
    2001, \aj, 121, 2308
  \bibitem[Romero et al.(1999)]{romero99} Romero, G.~E., Cellone,
    S.~A., \& Combi, J.~A.\ 1999, \aaps, 135, 477
  \bibitem[Santos-Lle{\'o} et al.(2001)]{santos01} Santos-Lle{\'o}, M.,
    et al.\ 2001, \aap, 369, 57
  \bibitem[Scargle(1981)]{scargle81} Scargle, J.~D.\ 1981, \apjs, 
    45, 1 
  \bibitem[Shakura \& Syunyaev(1973)]{shak73} Shakura, N.~I., \&
    Syunyaev, R.~A.\ 1973, \aap, 24, 337
  \bibitem[Shakura \& Sunyaev(1976)]{shak76} Shakura, N.~I., \&
    Sunyaev, R.~A.\ 1976, \mnras, 175, 613
  \bibitem[Shemmer et al.(2001)]{shemmer01} Shemmer, O., et al.\ 2001,
    \apj, 561, 162
  \bibitem[Siemiginowska \& Czerny(1989)]{aneta89} Siemiginowska, A.,
    \& Czerny, B.\ 1989, \mnras, 239, 289
  \bibitem[Spergel et al.(2003)]{wmap} Spergel, D.~N., et al.\ 2003,
    \apjs, 148, 175
  \bibitem[Stalin et al.(2004)]{stalin04} Stalin, C.~S.,
    Gopal-Krishna, Sagar, R., \& Wiita, P.~J.\ 2004, \mnras, 350, 175
  \bibitem[Stalin et al.(2005)]{stalin05} Stalin, C.~S., Gupta, A.~C.,
    Gopal-Krishna, Wiita, P.~J., \& Sagar, R.\ 2005, \mnras, 356, 607
  \bibitem[Starling et al.(2004)]{starling04} Starling, R.~L.~C.,
    Siemiginowska, A., Uttley, P., \& Soria, R.\ 2004, \mnras, 347, 67
  \bibitem[Stella \& Rosner(1984)]{stella84} Stella, L., \& Rosner,
    R.\ 1984, \apj, 277, 312
  \bibitem[Stirpe et al.(1994)]{stirpe94} Stirpe, G.~M., et al.\ 1994,
    \apj, 425, 609
  \bibitem[Svensson \& Zdziarski(1994)]{svensson94} Svensson, R., \&
    Zdziarski, A.~A.\ 1994, \apj, 436, 599
  \bibitem[Szuszkiewicz(1990)]{szusz90} Szuszkiewicz, E.\ 1990,
    \mnras, 244, 377
  \bibitem[Tr{\`e}vese et al.(2001)]{trevese01} Tr{\`e}vese, D., Kron,
    R.~G., \& Bunone, A.\ 2001, \apj, 551, 103
  \bibitem[Tr{\`e}vese \& Vagnetti(2002)]{trevese02} Tr{\`e}vese, D.,
    \& Vagnetti, F.\ 2002, \apj, 564, 624
  \bibitem[Turner(2004)]{turner04} Turner, N.~J.\ 2004, \apjl, 605,
    L45
  \bibitem[Ulrich et al.(1997)]{ulrich97} Ulrich,
    M.-H., Maraschi, L., \& Urry, C.~M.\ 1997, \araa, 35, 445
  \bibitem[Uttley et al.(2005)]{uttley05} Uttley, P., McHardy, 
    I.~M., \& Vaughan, S.\ 2005, \mnras, 359, 345 
  \bibitem[van der Klis(1997)]{klis97} van der Klis, M.\ 1997, 
    Statistical Challenges in Modern Astronomy II, 321 
  \bibitem[Vanden Berk et al.(2004)]{vanden04} Vanden Berk, D.~E.,
    et al.\ 2004, \apj, 601, 692
  \bibitem[Vasudevan \& Fabian(2007)]{vasud07} Vasudevan, R.~V., \&
    Fabian, A.~C.\ 2007, \mnras, 381, 1235
  \bibitem[Vestergaard \& Peterson(2006)]{vest06} Vestergaard, M., \&
    Peterson, B.~M.\ 2006, \apj, 641, 689
  \bibitem[Wilhite et al.(2008)]{wilhite08} Wilhite, B.~C., Brunner,
    R.~J., Grier, C.~J., Schneider, D.~P., \& vanden Berk, D.~E.\
    2008, \mnras, 383, 1232
  \bibitem[Wold et al.(2007)]{wold07} Wold, M., Brotherton, M.~S., \&
    Shang, Z.\ 2007, \mnras, 375, 989

\end{thebibliography}
\end{document}